\documentclass[preprint2]{aastex62}
\usepackage{graphicx}
\usepackage{subfigure}
\usepackage{natbib}
\usepackage{longtable}
\usepackage{rotating}
\graphicspath{{pic/}}
\usepackage{tikz}
\usetikzlibrary{shapes,arrows}
\usetikzlibrary{positioning}

\shorttitle{Catalogue of HT Galaxies}
\shortauthors{Pan et al.}

\begin{document}

\title{Catalog of One-side Head–Tail Galaxies in the FIRST Survey}

\correspondingauthor{Heng Yu}
\email{yuheng@bnu.edu.cn}

\author{Tong Pan}
\affiliation{Department of Astronomy, Beijing Normal University, Beijing, 100875, China.}
 
\author[0000-0001-8051-1465]{Heng Yu }
\affil{Department of Astronomy, Beijing Normal University, Beijing, 100875, China.}

\author[0000-0000-0000-0000]{Reinout J. van Weeren}
\affiliation{Leiden Observatory, Leiden University, PO Box 9513, 2300 RA Leiden, The Netherlands}

\author[0000-0000-0000-0000]{Shumei Jia }
\affiliation{Key Laboratory of Particle Astrophysics, Institute of High Energy Physics, Chinese Academy of Sciences, Beijing
100049, China}

\author[0000-0000-0000-0000]{Chengkui Li }
\affiliation{Key Laboratory of Particle Astrophysics, Institute of High Energy Physics, Chinese Academy of Sciences, Beijing
100049, China}

\author{Yipeng Lyu}
\affiliation{Department of Astronomy, Beijing Normal University, Beijing, 100875, China.}

\begin{abstract}

One-side head–tail (OHT) galaxies are radio galaxies with a peculiar shape. They usually appear in galaxy clusters,
but they have never been cataloged systematically. We design an automatic procedure to search for them in the
Faint Images of the Radio Sky at Twenty-Centimeters source catalog and compile a sample with 115 HT
candidates. After cross-checking with the Sloan Digital Sky Survey photometric data and catalogs of galaxy
clusters, we find that 69 of them are possible OHT galaxies. Most of them are close to the center of galaxy clusters.
The lengths of their tails do not correlate with the projection distance to the center of the nearest galaxy clusters,
but show weak anticorrelation with the cluster richness, and are inversely proportional to the radial velocity
differences between clusters and host galaxies. Our catalog provides a unique sample to study this special type of
radio galaxies. 

\end{abstract}

\section{Introduction} 
\label{sec:intro}
Radio galaxies are galaxies with high radio luminosity and relativistic particle emission. They have been studied for decades of years and show various morphologies\citep{1980ARA&A..18..165M,2011ApJS..194...31P,2020MNRAS.493.3811S}, which are closely related
to the activities of their central active galactic nucleus (AGN),
environment, and local galaxy richness\citep{2008LNP...740..143F,2017MNRAS.466.4346M}. The morphological
classification of radio galaxies could provide useful information
on the mechanism involved and act as a tracer of the
intragalatic environment.

\citet{2011ApJS..194...31P} systematically examined the Faint Images of
the Radio Sky at Twenty-Centimeters \citep[FIRST;][]{1995ApJ...450..559B} catalog and assorted sources by the number of member
components. They presented 15 types of radio sources,
including wide-angle tail (WAT), narrow-angle tail (NAT),
core–jet, W-shaped, X-shaped, and so on.

The Radio Galaxy Zoo project \citep{2015MNRAS.453.2326B} provides a different approach. It recruited thousands of
volunteers to do a visual inspection of the host galaxy and
radio morphology with the FIRST and optical images \citep{2019AJ....157..126G}.

There is a special type of radio galaxy—the head–tail (HT)
—which has not been systematically searched for. The first HT
was discovered in the Perseus cluster \citep{1968MNRAS.138....1R}. \citet{1972Natur.237..269M} nominated these remarkable radio galaxies
as HT galaxies and interpreted them as radio trails along
trajectories through a dense intergalactic medium. Because
WATs and NATs have similar bright heads and long faint tails,
they are also sometimes mentioned as HT galaxies \citep[e.g.][]{2009MNRAS.392.1070M,2013MNRAS.432..243P}. 

This paper will focus on the one-sided head–tail (hereafter
OHT) galaxies. Unlike WATs or NATs, they have only one
unresolved tail. Some of them have been resolved as NATs
with high-resolution observations \citep{2017A&A...608A..58T}. so they are also called narrow HT \citep{2017A&A...608A..58T}, and sometimes head-tailed galaxies for simplicity \citep[e.g.][]{1996MNRAS.282..137J,2018ApJ...853..100Y,2020MNRAS.493.3811S}.
In this paper, we use the term HT to represent all
three types: NAT, WAT, and OHT.

It is still unclear how the tails of OHTs formed. A possible
explanation for the formation of such a structure may be the
existence of a massive structure such as a galaxy cluster that
makes the HT galaxy infall with a high velocity, resulting in the
merger of two radiation lobes along the opposite direction of
motion\citep{2019ApJ...887...26O}. 

However, most known OHTs are by-products of cluster
studies. This fact may bias our understanding of them. To
explore their situation in a more general way, we need a fair
sample. The FIRST project based on the Very Large Array
(VLA) is a useful radio source catalog. It is a project designed
to produce the radio equivalent of the Palomar Observatory
Sky Survey over 10,000 square degrees of the north and south
Galactic caps. Compared with other available radio sky surveys
such as the NRAO/VLA Sky Survey\citep{1998AJ....115.1693C} and
the TIFR GMRT Sky Survey\citep{2017A&A...598A..78I}, FIRST has a
much higher resolution, which is crucial to our study. The
ongoing LOFAR Two-meter Sky Survey \citep[LoTSS;][]{2017A&A...598A.104S,2019A&A...622A...1S} will provide more helpful data in the near
future.

This paper is organized as follows. In Section 2, we describe
the preliminary selection criteria of HT-like structures in
FIRST. Their optical identification is presented in Section 3.
Section 4 concerns cluster checking. Properties of this sample
are presented in Section 5. Section 6 contains our summary and
conclusions.

\section{radio identification} 
\label{sec:style}
The FIRST Survey used the VLA in its B configuration
centered at 1.4 GHz from 1993 through 2011 and acquired 3
minutes snapshots covering about 10,575 square degrees of sky\citep{1995ApJ...450..559B}. Its latest source catalog, which was
released on 2014 December 17, contains 946,432 sources.
They are derived from fitting the flux density of each source with an elliptical Gaussian\citep{1997ApJ...475..479W}. The peak,
integrated flux densities, and sizes in the catalog provide a good
representation of the source morphology.

Considering the particular shape of OHT sources, we design
a straightforward routine to identify them from the radio
catalog. We start with elongated sources which look like a
“tail”, and search for bright sources, which could be a “head”,
on the tip of them. If their relative positions, alignments, and
brightness satisfy the given criteria, we combine them as HT
structures.

\subsection{``Tail''}
The lowest signal-to-noise ratio (S/N) of FIRST sources is
5. To get rid of ambiguous radio sources in the catalog, we
focus on sources with a S/N larger than 10. Since the angular
resolution of FIRST is 5$^{\prime\prime}$, we only care about resolved sources
and constrain the half major and minor axes (fMaj and fMin)
derived from the elliptical Gaussian model larger than 5$^{\prime\prime}$.
There are 453,346 sources satisfying these two basic conditions
in the catalog.

\begin{figure}
\includegraphics[width=0.45\textwidth]{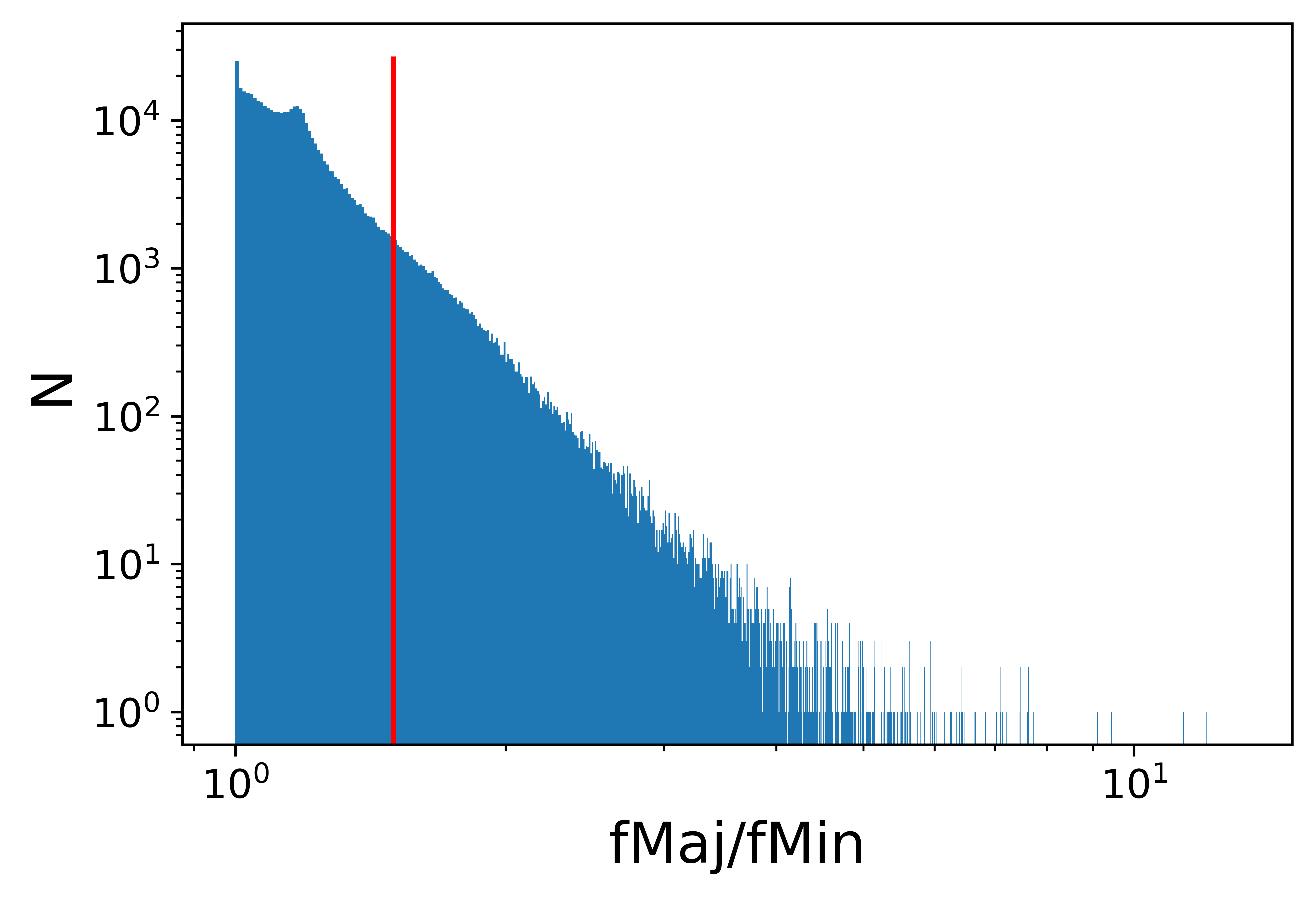}\\
\caption{The ellipticity distribution of 453,346 FIRST sources. The red line refers to the ellipticity \textit{e} = 1.5.}
\label{fig:hist1}
\end{figure}

The tail of an OHT source is always elongated, while the
beam shape of the FIRST survey is round. So the ratio between
the major and minor axes (ellipticity, \textit{e} = fMaj/fMin) is the
main difference between point sources and extended sources in
the FIRST catalog. The ellipticity histogram of all 453,346
sources is given in Fig.\ref{fig:hist1}. Considering tails could be short
due to the projection effect, we adopt a conservative ellipticity
value \textit{e} $>$ 1.5 as our tail selection criterion to include more
possible candidates. With this criterion, 49,086 “tail” sources
are sifted out.

\subsection{``Head''}

\begin{figure}
\includegraphics[width=0.45\textwidth]{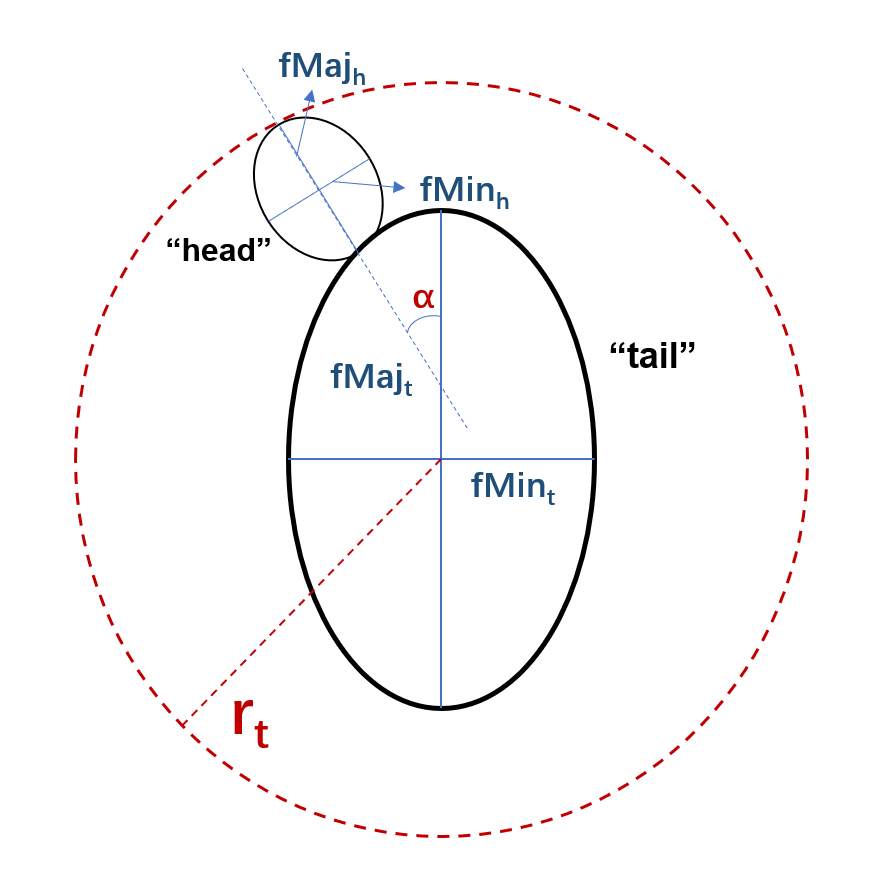}\\
\caption{The red dashed circle r$_t$ refers to the range of 1.5 times the half major axes of the tail (fMaj$_t$); within this range, we identify the brightest source as the ``head''. $\alpha$ in red is the angle between the long axis of the tail and the head,
which is the phase-angle offset used to determine whether the “head” and “tail”
are aligned.}
\label{fig:show2}
\end{figure}

All known OHT sources show bright radio cores close to the
host galaxy, if not overlapping with it. The core is called the
“head”. So we go through all tail candidates to check if there is
any bright source around them. On one hand, the head’s peak
flux should be brighter than the value of the tail: fPeak$_h$ $>$ fPeak$_t$.
On the other hand, the ellipticity of the head
should be smaller than the tail:
$e_h$ $<$ $e_t$.
Considering that some
heads may not have close contact with the tail, we set a slightly
large searching radius \textit{r}$_t$ = 1.5 fMaj$_t$, as Fig. \ref{fig:show2} illustrates.The
brightest source within this radius is chosen as the “head”
associated with the “tail”. There are 9704 tails with a head
beside a tail.

To remove bright sources overlapping with tails by chance, we add an additional condition -- the phase-angle offset $\alpha$ between the tail and the head (shown in Fig.\ref{fig:show2}) should be less than 20$^\circ$ to guarantee their alignment, because both of
them follow the trajectories of the host galaxy.

\begin{figure}
\includegraphics[width=0.45\textwidth]{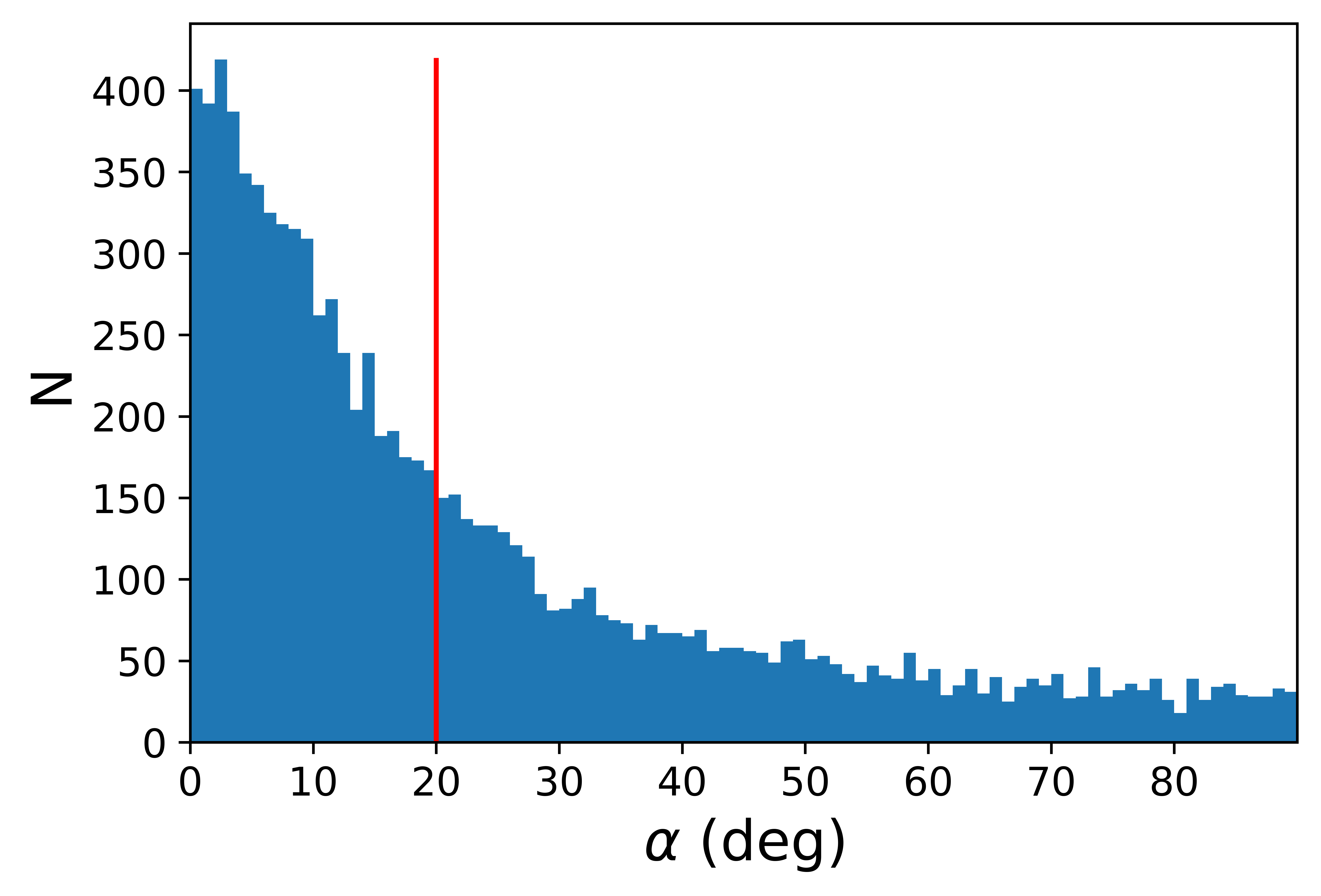}\\
\caption{The histogram distribution of the phase-angle offset $\alpha$ of 9704 "tails" with a "head". The red solid line refers to $\alpha = 20^\circ$.}
\label{fig:hist2}
\end{figure}

\begin{figure}[htp]
 \centering
 \includegraphics[width=0.48\textwidth]{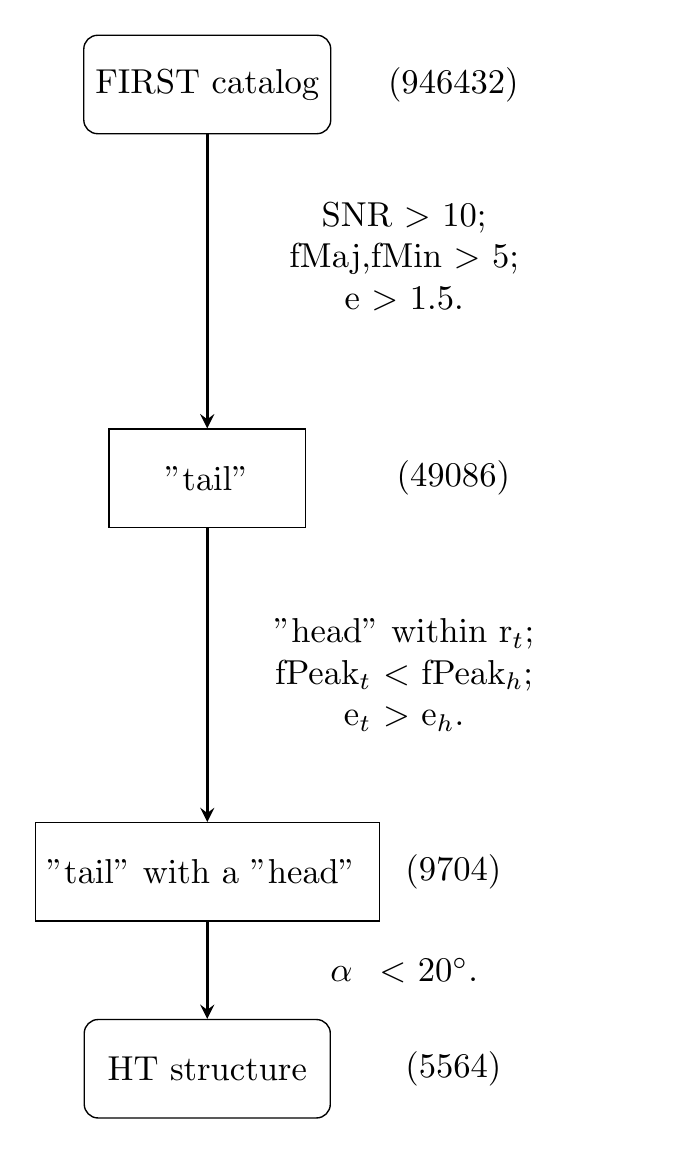}
\caption{ Flow chart of the procedure for the radio image identification. The total number in each catalog is labeled alongside its rectangle. The selection conditions are listed between the steps. }
\label{fig:fc1}
\end{figure}

The histogram distribution of $\alpha$ is given by Fig.\ref{fig:hist2}. It is clear that $\alpha<20 ^\circ$ is not strict.
A flow chart (Fig.\ref{fig:fc1}) is given to show our selection process more intuitively.
After the
head searching, we get 5564 head–tail combinations as our HT
structures.

\begin{figure}
\includegraphics[width=0.45\textwidth]{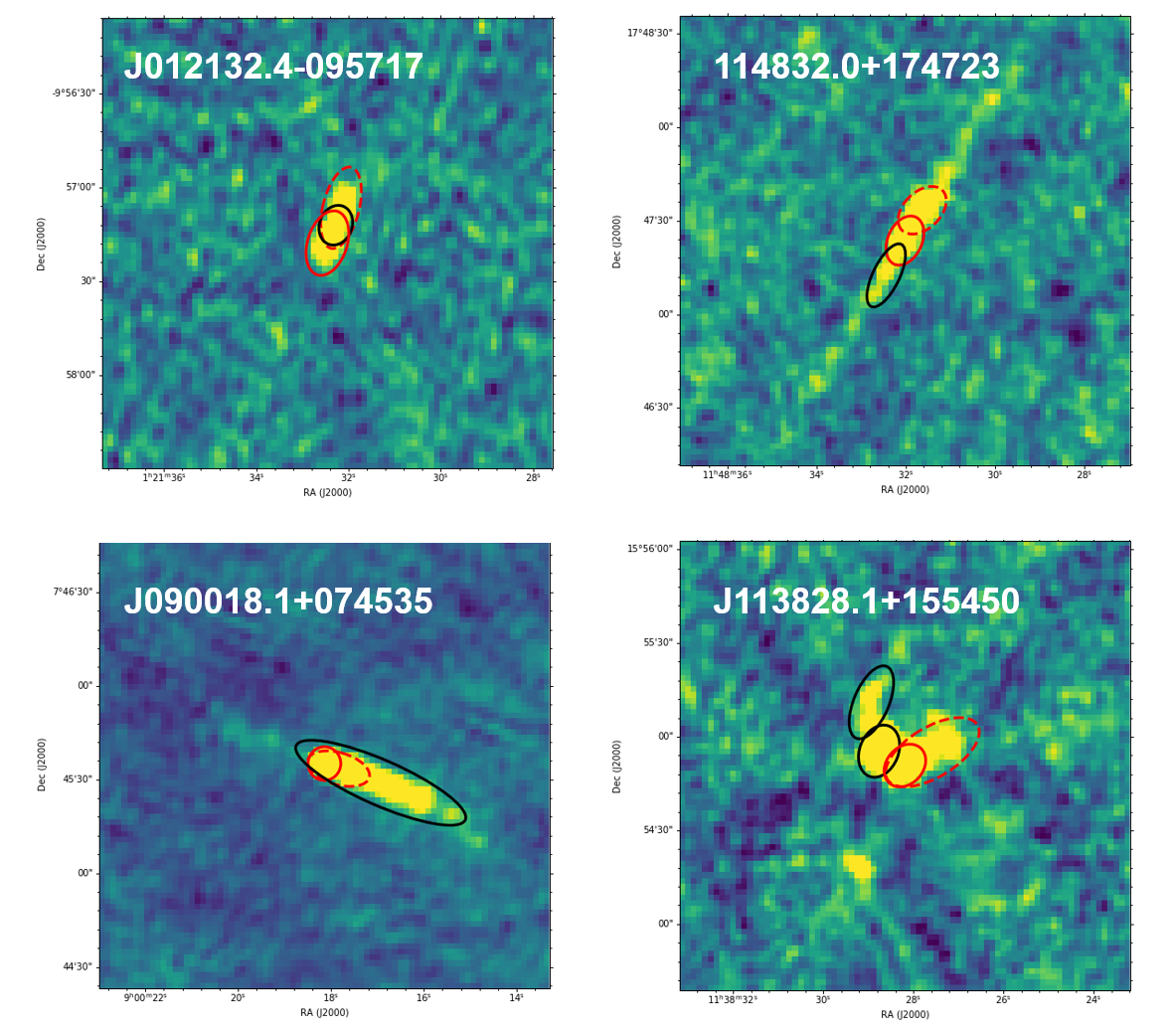}\\
\caption{This figure shows four typical spurious sources showing HT-like patterns. All panels are show the same area of 2.4$^\prime$ $\times$ 2.4$^\prime$ and ellipses are Gaussian fittings from the FIRST catalog. Red solid ellipses refer to a ``head'' and red dashed ellipses mean ``tail''. The white text label of each image is the FIRST ID of the ``head" in the center. All these four structures satisfy our selection criteria, but they are not real OHT sources.}
\label{fig:case1}
\end{figure}

However, there are still many spurious sources showing HT-like patterns, e.g. parts of wide-angle tails, radio jets, small lobes, or even calibration artifacts. 
Fig.\ref{fig:case1} provides four cases to show this variety.  J012132.4-095717 is an AGN with double jets because its optical counterparts (which will be checked later) are located in the middle of its head and tail. J114832.0+174723 passes our selection because of the contamination of calibration artifacts. J090018.1+074535 is part of a giant radio jet and there is a lobe on the opposite side, outside this figure. J113828.1+155450 is a part of a wide-angle tail structure.

So it is difficult to identify OHTs with radio images alone.
To select more reasonable candidates, we take optical images into account.

\section{Optical Counterpart} 
\label{subsec:tables}
The head of a real HT source coincides with its host galaxy,
while the shape of the tail shows its trajectory on the sky plane.
We cross-check our HT structures with the Sloan Digital Sky
Survey (SDSS) data. We adopt the photometric catalog of
SDSS DR12 \citep{2015ApJS..219...12A}. 
It covers 68\% of the northern sky and about 90\% of the FIRST area.

\begin{figure}[htp] 
 \centering
 \includegraphics[width=0.48\textwidth]{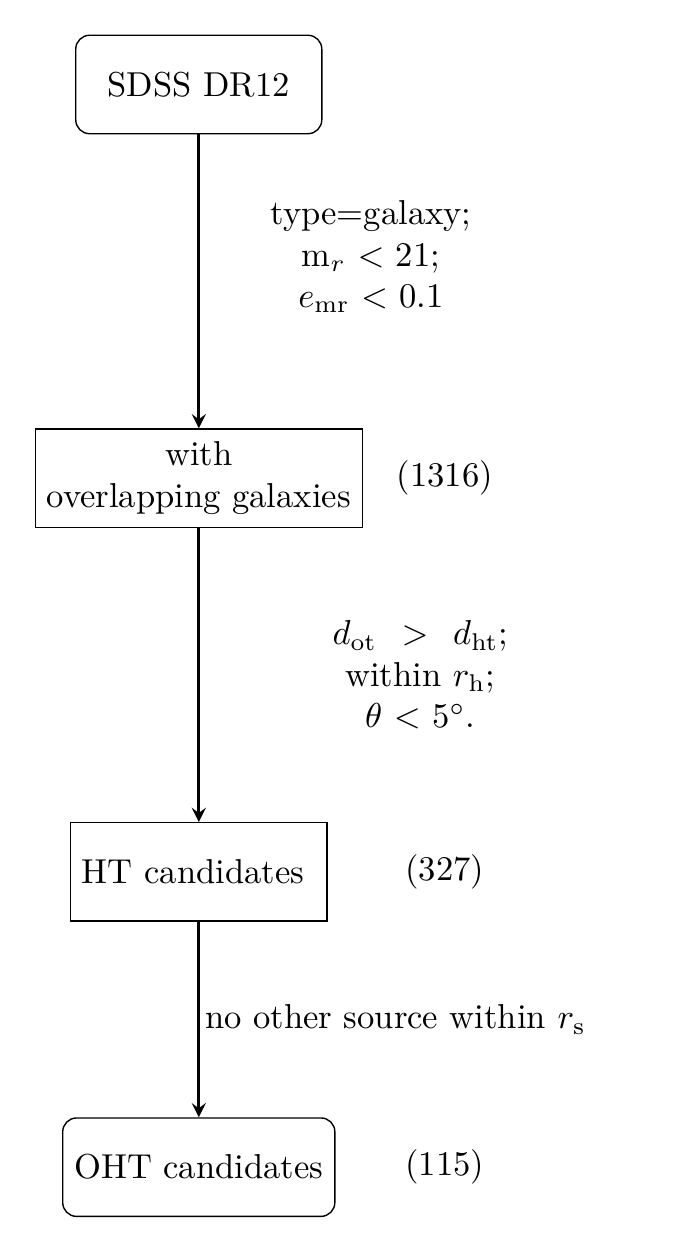}
\caption{ Flow chart of the procedure for the SDSS cross-checking. The total number of each catalog is labeled alongside its rectangle. The selection conditions are listed between the steps.} 
\label{fig:fc2}
\end{figure}

For each HT structure, we search for galaxies around its head in the SDSS photometric catalog. The automatic procedure of this section is illustrated by the flowchart Fig.\ref{fig:fc2}.
There are 90\% of 5564 HT structures that have optical objects within 3$^\prime$, but many of them are faint sources. To avoid unreliable faint sources, we only consider galaxies that are bright enough, that is to say, they have an \textit{r}-band magnitude \textit{m}$_r$ $<$ 21 mag with an error $e_{mr}$ $<$ 0.1 mag. We then get 1316 galaxies corresponding with our HT structures. 

\begin{figure}
\includegraphics[width=0.45\textwidth]{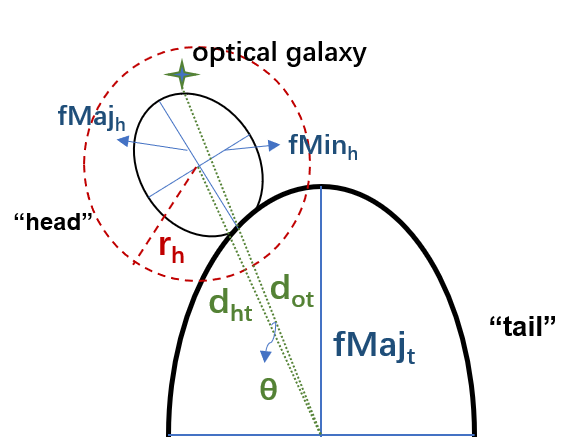}\\
\caption{The red dashed circle \textit{r}$_h$ refers to the radius \textit{r}$_h$ = 1.5 times of the half major axis. The green star refers to the galaxy near the head. The two green dotted lines are the distance of head and tail with the galaxy and tail. $\theta$ is the angle between the two directions which must less than 5$^\circ$ to make sure the galaxy aligns with the system.}
\label{fig:show3}
\end{figure}

\begin{figure}
\includegraphics[width=0.45\textwidth]{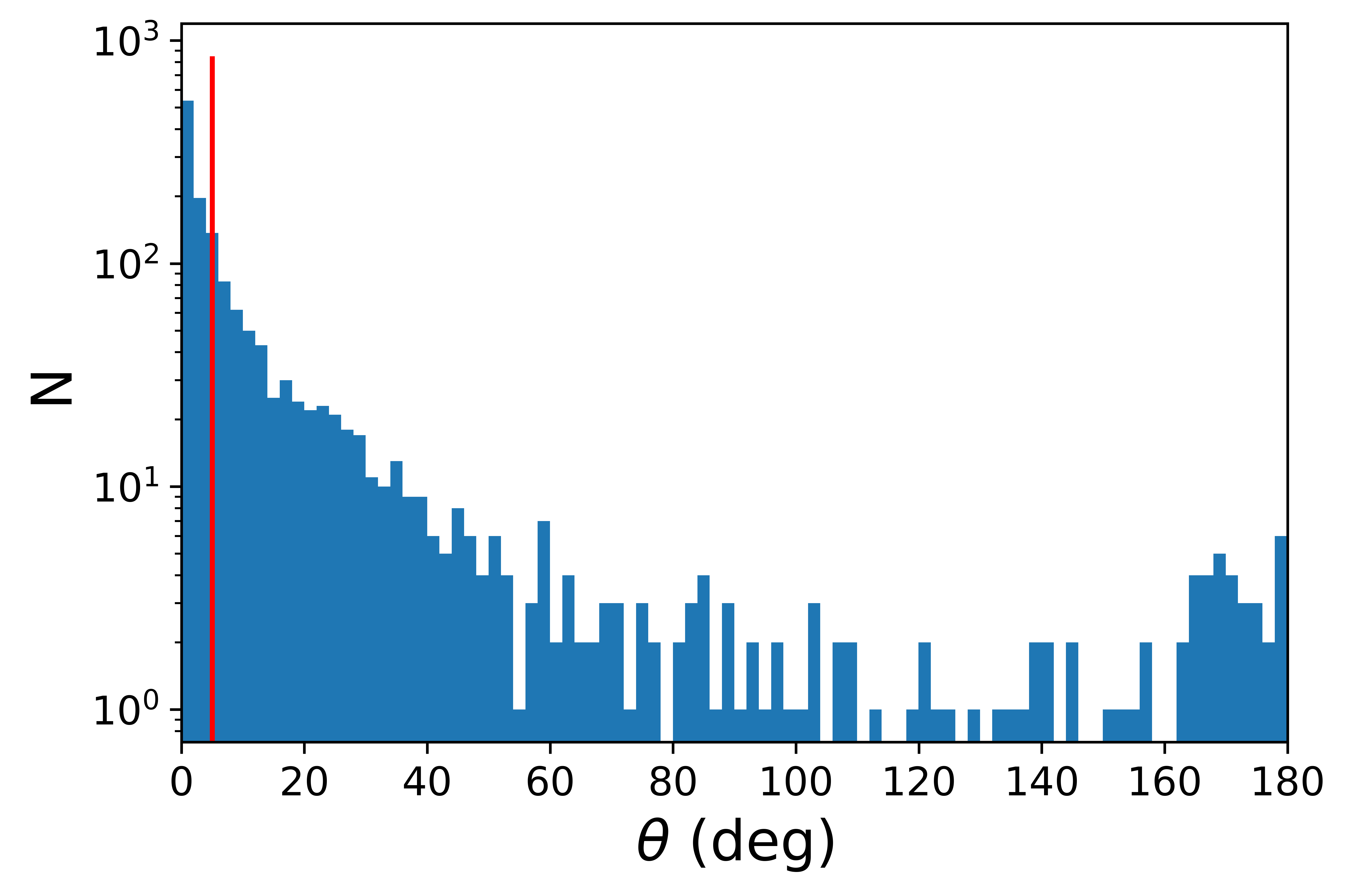}\\
\caption{The histogram distribution of $\theta$ of 1316 galaxies and the red line refers to $\theta$ = 5$^{\circ}$.}
\label{fig:histtheta}
\end{figure}

 Next, we select galaxies within a radius \textit{r}$_h$ = 1.5 times of the half major axis of the head (fMaj$_h$), and choose the closest one as the optical counterpart. 
In some cases, an optical galaxy is
just located between the head and tail, as in the first case of Fig.\ref{fig:case1}.
This means both the head and the tail are actually jets
of the galaxy.
To filter off these cases, we add additional criteria:
the distance between the optical galaxy and the tail (\textit{d}$_{ot}$) should be larger than the distance between the head and the tail (\textit{d}$_{ht}$), and the angle between the two lines $\theta$ should be less than 5 $^\circ$. So the optical galaxy will appear on the far end of
the HT structure and align with the moving direction of the
system. This illustration is given by Fig.\ref{fig:show3}. 
The histogram distribution of $\theta$ is shown in Fig.\ref{fig:histtheta}. 

\begin{figure}
\includegraphics[width=0.45\textwidth]{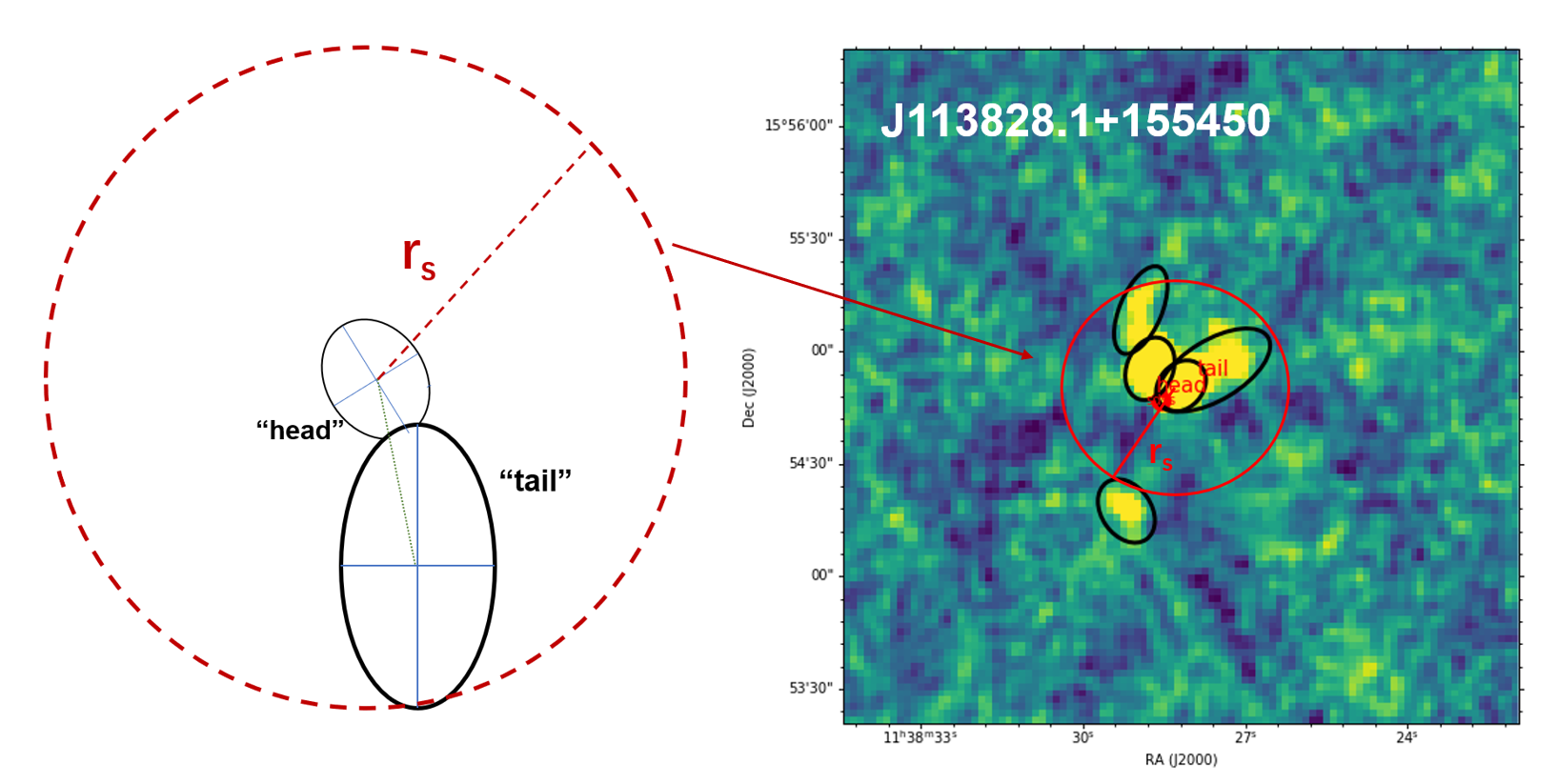}\\
\caption{The red circle refers to the radius of the HT size \textit{r}$_s$. We require there is no other source within \textit{r}$_s$. J113828.1+155450 is a typical WAT. A part of it is selected as an OHT candidate. It is rejected by the isolation criterion.}
\label{fig:show4}
\end{figure}

There are 327 galaxies satisfying all the above criteria.
However, some of them are still a part of a double-jet radio
galaxy, WAT galaxy, or NAT galaxy. We further add an
isolation criterion: if there are any other radio sources except
the tail and the head in the radius of the HT size $ r_s = d_{ht} + fMaj_{t}$, we exclude the candidate. With this isolation criterion, we reduce our sample size to 115. The illustration is given by Fig.\ref{fig:show4}.    

Since the jets of some radio sources are separated a lot on the
sky plane, beyond our checking radius r$_s$, there is no effective
way to automatically identify them. We visually check our
sample, and recognize another 21 jet components, 2 mergers,
and 2 irregular patterns, and remove them. There are 90
isolated OHT candidates remaining in our catalog.

Additionally, there are three known OHT in the FIRST fields not recognized by our procedure. Both J134859.3+263334 in A1795 and J115508.8+232615 in A1413\citep{2019A&A...622A..24S} are fitted with a single elongated Gaussian component. The third OHT is J134150.5+262217 in A1775\citep{2017A&A...608A..58T}. It is rejected because of a nearby radio source belonging to the brightest central galaxy (BCG) of its host cluster. There are another two candidates found by chance—J084115.3
+075809 and J000313.1-060712. They are not selected by our
procedure but noticed at the visual check stage. The former has
a faint tail with a S/N of 7.6, smaller than our threshold of 10,
while the latter is fitted with one Gaussian component. We add
these five cases to our catalog manually, so the total number of
candidates reaches 95.

\begin{figure}
\includegraphics[width=0.45\textwidth]{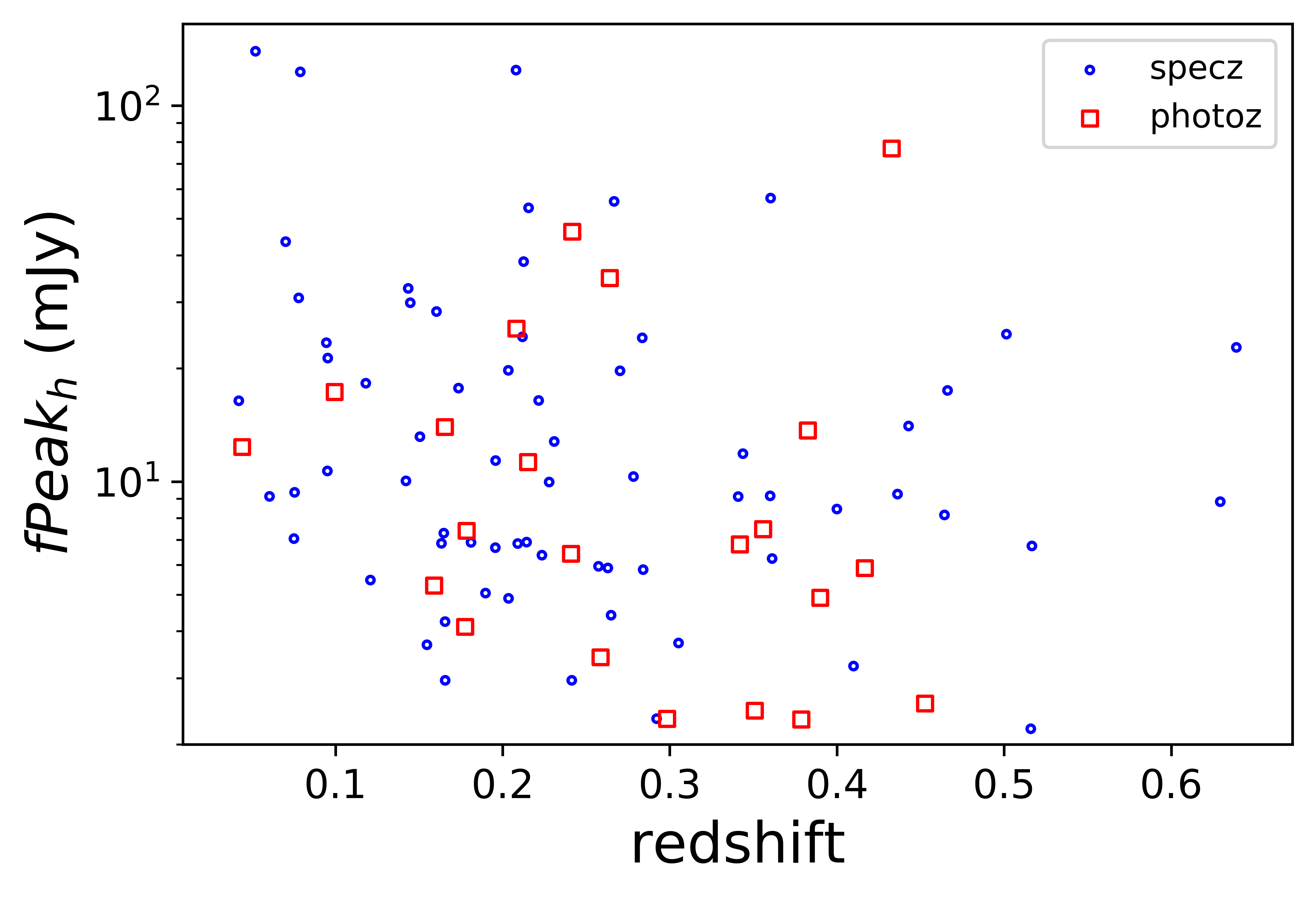}\\
\caption{The peak fluxes of head and redshift distribution. Blue dots refer to galaxies with photometric redshifts while red dots refer to galaxies with spectroscopic redshifts.}
\label{fig:histpeak}
\end{figure}

There are 71 that have optical counterparts with spectroscopic measurements. 
For the rest, we adopt photometric redshifts \citep{2015ApJS..219...12A}.
The peak fluxes and redshift distribution of these candidates are shown in Fig.\ref{fig:histpeak}.

To estimate possible missing candidates likes the five
sources mentioned above, we visually check 1000 fields
selected randomly from 49,086 tail candidates. There are 138
radio HT structures found. The total number of HT structures
can then reach to up 6774. Considering that only 115 out of
5564 HT structures pass all OHT checks, there will be about
140 OHT candidates that remain in the end. So we conclude
that there might be about 18\% (25 out of 140) OHT candidates
that are missing when using our procedure.

\section{cluster association}
With some well-studied cases \citep{2017A&A...608A..58T}, OHT structures are resolved as galaxies with high peculiar
velocities located in the central region of clusters. Their two radio jets are bent as one along the opposite direction of their movement. But this scenario has only been verified in in few cases. With this new sample, we could verify this interpretation in a more general way.

We cross-check our OHT candidates with the Abell catalog \citep{1989ApJS...70....1A} \footnote{We adopt the centers and redshifts given by the NED database instead of values in the catalog.} and cluster catalogs derived from the SDSS survey, e.g. the WHL2015 catalog \citep{2015ApJ...807..178W}, the MSPM catalog \citep{2012MNRAS.422...25S}, the SDSSCG catalog \citep{2009MNRAS.395..255M,2011MNRAS.418.1409M}, catalogs contained in the NASA/IPAC Extragalactic
Database\citep[NED;][]{2017IAUS..325..379M} to search for possible host clusters around OHT candidates.

We check galaxy clusters within 1 $^\circ$ of each OHT candidate (\textit{r}$_s$ $<$ 1$^\circ$ ). This radius corresponds to a projection distance of 6.7 Mpc at redshift 0.1, and 19.5 Mpc at redshift 0.4. We choose the closest cluster with a redshift difference smaller than 0.02 \citep{2016MNRAS.460.1371B} as the association cluster. We then find association clusters for 89 out of 95 OHT candidates. All of the six sources without cluster association have relatively faint optical counterparts.

\begin{figure}
\includegraphics[width=0.45\textwidth]{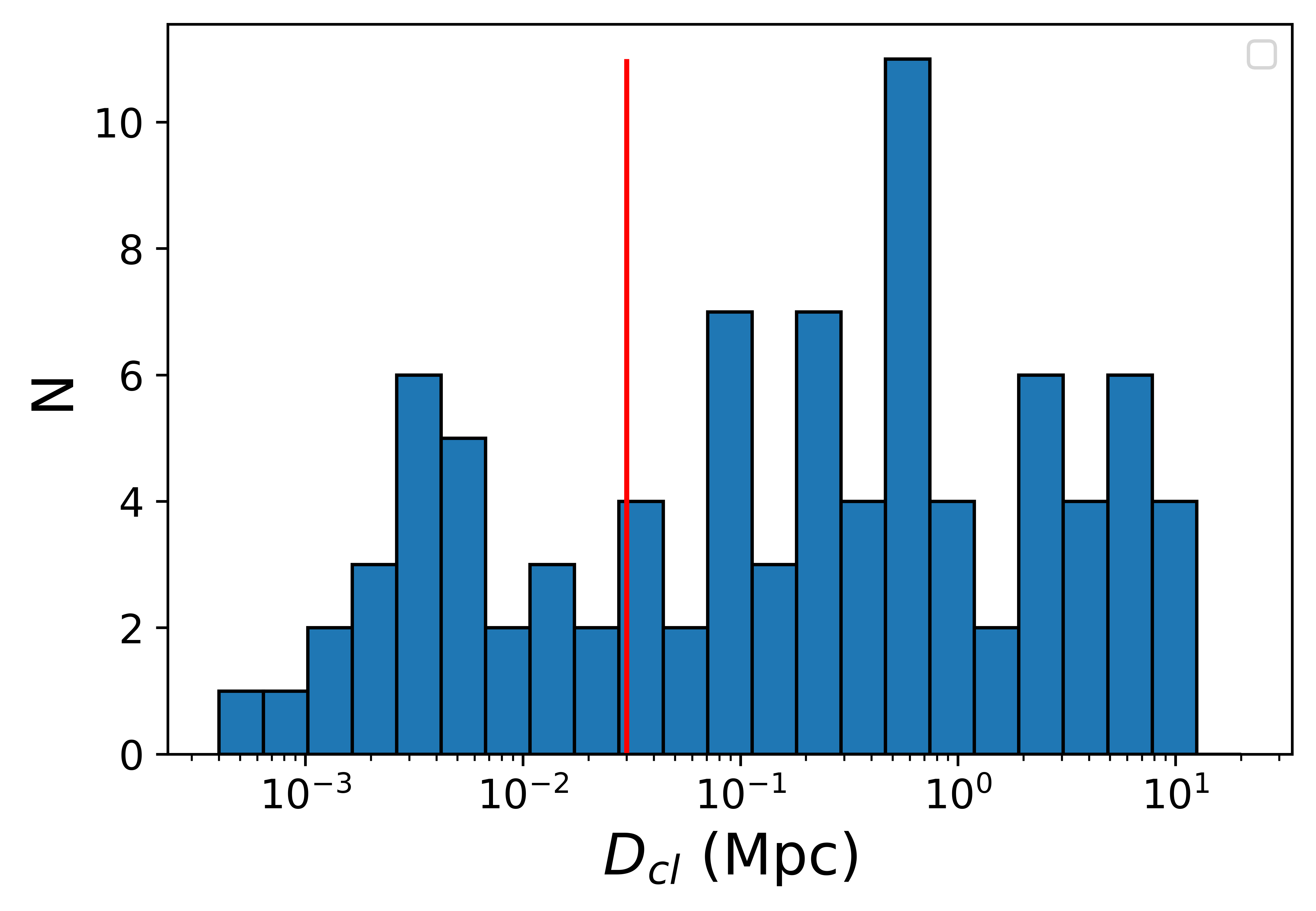}\\
\caption{The distribution of projected distances between OHT candidates who have SDSS galaxy counterparts and the nearby galaxy clusters. The red solid line refers to 30 kpc, which is the typical size of elliptical galaxies\citep{2010MNRAS.409.1362D}.}
\label{htsep2}
\end{figure}

The histogram of the projected distance between isolated OHT candidates and association clusters is given by Fig.\ref{htsep2}. Sixty-five OHT candidates appear within 1 Mpc of association
clusters. Twenty-six of them are located within a projected
distance of 30 kpc, which is the typical size of elliptical
galaxies\citep{2010MNRAS.409.1362D}. Their host galaxies are more likely to
be the central dominant (cD) galaxies of clusters. Since the cD
galaxies could not have a large peculiar velocity, their tail-like
structures might be parts of double jets. The opposite
component on the other side is invisible due to the relativistic
beaming effect\citep[e.g.][]{1992Natur.355..804S,2002MNRAS.336..328L}, so we remove them and get 69 OHT galaxies in the end.

\section{Sample properties} \label{sec:style}

The basic properties of our OHT sample are listed in Table \ref{tab:1}.
The first column is the serial number in our sample, the second
column is the source id in the FIRST catalog, followed by the
SDSS ID of its optical counterpart, the r-band magnitude, and
the redshift. The sixth to eighth columns are the name of the
association cluster, its redshift, and the projection distance
between the head and the cluster center. There are six OHT
candidates without cluster associations. Twenty-one host
galaxies of OHT candidates have only photometric redshifts.

\LTcapwidth=\textwidth
\begin{longtable*}{ccccccc c}

\hline
       & FIRST ID & SDSS ID & \textit{m}$_r$ & \textit{z}$_g$ & Cluster & \textit{z}$_{cl}$ & \textit{D}$_{cl}$ \\
       &        &         &         &     &         &    & (Mpc) \\ 
\hline
    1 & J000323.0-060458 & 1237672793959891057 & 17.62  & 0.241* & A2697 & 0.234  & 0.657  \\
    2 & J004857.6+115641 & 1237678858476847541 & 19.68  & 0.390* & WHL J004906.0+115754 & 0.409*  & 0.790  \\
    3 & J020159.9+034343 & 1237678660900421942 & 17.07  & 0.166 & A293 & 0.165  & 0.217  \\
    4 & J023834.4-032910 & 1237679255210623312 & 19.52  & 0.298* & WHL J023949.1-033022 & 0.317  & 5.245  \\
    5 & J030013.1-051514 & 1237679439892643986 & 18.14  & 0.264* & WHL J030100.9-051223 & 0.266*  & 3.041  \\
    6 & J035820.9+004223 & 1237666301633364230 & 19.30  & 0.383* & WHL J035820.6+003829 & 0.393*  & 1.266  \\
    7 & J071130.2+390729 & 1237673429620032132 & 20.12  & 0.453* &   &   &  \\
    8 & J075357.7+420255 & 1237651192426004963 & 19.36  & 0.351* & WHL J075400.8+420246 & 0.367  & 0.183  \\
    9 & J075431.9+164822 & 1237664835922100518 & 14.77  & 0.044* & MSPM 668 &  0.046* & 0.033  \\
    10 & J081859.7+494635 & 1237651272422522889 & 16.01  & 0.095 & WHL J082041.5+492231 & 0.077  & 2.60  \\
    11 & J085146.0+371440 & 1237657627901559004 & 18.31  & 0.178* & WHL J085146.2+371416 & 0.172  & 0.073  \\
    12 & J085732.8+592751 & 1237663546906050701 & 17.28  & 0.203 & WHL J085748.0+592925 & 0.203  & 0.502  \\
    13 & J090327.1+042614 & 1237658423006396511 & 18.23  & 0.361 & WHL J090429.6+040433 & 0.362  & 8.197  \\
    14 & J091035.7+350741 & 1237664871895794181 & 19.84  & 0.516 & WHL J091028.8+350836 & 0.519  & 0.641  \\
    15 & J091327.7+555823 & 1237651191360586025 & 18.08  & 0.259* & WHL J091332.3+555857 & 0.269  & 0.211  \\
    16 & J094613.8+022246 & 1237653665257357349 & 16.41  & 0.118 & A869 & 0.120  & 0.105  \\
    17 & J100623.5+240526 & 1237667293731684482 & 15.49  & 0.075 & MSPM 6798 &  0.075* & 0.471  \\
    18 & J100850.4+135538 & 1237671260133065021 & 16.78  & 0.204 & WHL J100840.4+135750 & 0.201  & 0.660  \\
    19 & J102102.9+470055 & 1237658614124314722 & 17.35  & 0.181 & WHL J102104.2+470054 & 0.179  & 0.041  \\
    20 & J102604.0+390523 & 1237661138497896734 & 16.98  & 0.145 & WHL J102622.9+390852 & 0.149  & 0.792  \\
    21 & J104506.2+083718 & 1237671930671006606 & 20.79  & 0.639 &   &   &  \\
    22 & J105624.7+164429 & 1237668585967714469 & 15.41  & 0.095 & SDSSCGB 22800 & 0.095  & 0.074  \\
    23 & J110532.0+073730 & 1237661972251213959 & 16.71  & 0.155 & WHL J110527.0+073836 & 0.154  & 0.268  \\
    24 & J111911.1+081538 & 1237661972789592202 & 15.02  & 0.076 & MSPM 5279 &  0.075* & 0.449  \\
    25 & J112540.2+333924 & 1237665024363266295 & 19.06  & 0.356* &  &   &   \\
    26 & J114357.6+510236 & 1237657627916108093 & 20.56  & 0.501 & WHL J114235.0+510545 & 0.501  & 4.967  \\
    27 & J115716.8+333629 & 1237665126931234949 & 18.06  & 0.214 & A1423 & 0.216  & 0.038  \\
    28 & J122643.5+195050 & 1237667915421057112 & 16.97  & 0.224 & WHL J122642.5+195026 & 0.222  & 0.104  \\
    29 & J122902.4+473655 & 1237661357545095254 & 17.68  & 0.263 & A1550 & 0.259  & 0.100  \\
    30 & J123449.2+031136 & 1237651737371410669 & 18.87  & 0.410 & WHL J123458.4+030449 & 0.409  & 2.371  \\
    31 & J123547.4+030301 & 1237651754551083230 & 18.29  & 0.284 & SDSSCGB 16587  & 0.284   & 0.045  \\
    32 & J124042.3+020822 & 1237651753477931371 & 18.99  & 0.178* & XMMXCS J1243.0+0233 & 0.19  & 8.354  \\
    33 & J124135.9+162033 & 1237668588651610193 & 15.04  & 0.070 & MSPM 2617 &  0.071* & 1.010  \\
    34 & J125017.5+084215 & 1237658491211022497 & 18.61  & 0.342* & WHL J125019.9+084209 & 0.341  & 0.176  \\
    35 & J125908.6+412937 & 1237662193992859752 & 17.89  & 0.278 & WHL J125900.0+413128 & 0.277  & 0.629  \\
    36 & J131525.2+171745 & 1237668624628842730 & 20.13  & 0.629 &   &   &  \\
    37 & J132418.5+373531 & 1237664846122582252 & 17.74  & 0.241 & WHL J132412.4+373334 & 0.240  & 0.533  \\
    38 & J134436.7+534422 & 1237658802580226295 & 17.13  & 0.166 & WHL J134456.3+534504 & 0.166  & 0.511  \\
    39 & J135521.7-025453 & 1237655498129867111 & 19.12  & 0.165* & WHL J135522.7-031944 & 0.171*  & 5.02   \\
    40 & J140412.9+562232 & 1237659144558281082 & 20.61  & 0.159* &   &   &  \\
    41 & J143258.5+291926 & 1237665101138952294 & 17.54  & 0.222 & WHL J143321.8+292701 & 0.220  & 1.964  \\
    42 & J144348.3-002601 & 1237648720711778680 & 18.29  & 0.305 & WHL J144335.3-002106 & 0.292  & 1.569  \\
    43 & J150746.8+040233 & 1237654880742277346 & 17.54  & 0.163 & WHL J150813.6+042405 & 0.162  & 3.814  \\
    44 & J150836.8+235224 & 1237665429713191101 & 17.01  & 0.196 & WHL J150747.3+234533 & 0.196  & 2.615  \\
    45 & J152045.1+483923 & 1237659163343716378 & 15.95  & 0.078 & WHL J152052.2+483938 & 0.075  & 0.103  \\
    46 & J152122.5+042030 & 1237662266464600297 & 13.86  & 0.052 & MSPM 2944 & 0.052* & 0.656   \\
    47 & J153845.3-014018 & 1237655498678010865 & 19.94  & 0.417* & WHL J153845.6-013914 & 0.409*  & 0.357  \\
    48 & J155618.5+213516 & 1237665127491567870 & 17.71  & 0.196 & WHL J155615.3+213506 & 0.197  & 0.151  \\
    49 & J155813.8+271621 & 1237662340012573171 & 16.37  & 0.095 & A2142 & 0.091  & 0.278  \\
    50 & J160149.7+490645 & 1237655349965750765 & 19.80  & 0.433* & MSPM 1271 &  0.432* & 4.949  \\
    51 & J161706.3+410646 & 1237665356696322197 & 17.73  & 0.267 & WHL J161658.3+412836 & 0.285  & 5.713  \\
    52 & J163802.7+162304 & 1237665231061648153 & 20.12  & 0.208* & WHL J163825.9+164018 & 0.181  & 3.348  \\
    53 & J164021.9+464246 & 1237651715872325877 & 18.84  & 0.208 & A2219 & 0.225  & 0.084  \\
    54 & J164058.8+114404 & 1237665567159484554 & 14.96  & 0.078 & WHL J163925.4+115131 & 0.085  & 2.329  \\
    55 & J165250.6+630029 & 1237671767467754423 & 20.30  & 0.379* & WHL J165504.5+623409 & 0.372*  & 9.511 \\
    56 & J172208.6+330640 & 1237665569838399705 & 15.83  & 0.099* & WHL J172216.4+330427 & 0.111  & 0.337  \\
    57 & J210451.0+050320 & 1237669762788557036 & 18.54  & 0.242* & WHL J210322.5+051633 & 0.258*  & 6.242  \\
    58 & J213319.1-063802 & 1237652936183185672 & 15.98  & 0.160 & SDSSCGB 21196  & 0.160  & 3.071 \\
    59 & J221807.6+085334 & 1237679009851441333 & 16.57  & 0.215* & WHL J221750.3+085544 &  0.234* & 1.085  \\
    60 & J224819.4-003641 & 1237666407360823478 & 17.33  & 0.213 & WHL J224659.4-012817 & 0.206  & 11.338  \\
    61 & J224830.6+122141 & 1237678860074615269 & 20.76  & 0.517 &   &   &  \\
    62 & J231953.0-011625 & 1237653010821480629 & 17.84  & 0.284 & WHL J231945.0-011729 & 0.287  & 0.593  \\
    63 & J232437.2+143833 & 1237656242242060352 & 15.78  & 0.042 & WHL J232420.1+143850 & 0.041  & 0.205  \\
    64 & J232844.5+000134 & 1237663784193949861 & 17.48  & 0.292 & WHL J232809.2+001109 & 0.278  & 3.342  \\
    65 & J134859.3+263334 & 1237665532782838032 & 15.75  & 0.061  & A1795 & 0.062  & 0.189  \\
    66 & J115508.8+232615 & 1237667736653857373 & 20.64  & 0.194* & A1413 & 0.143  & 0.440  \\
    67 & J134150.5+262217 & 1237665532245311533 & 14.42  & 0.069  & A1775 & 0.072  & 0.060  \\
    68 & J084115.3+075809 & 1237661066018488762 & 18.52  & 0.220* & WHL J084114.3+075844 & 0.245  & 0.146  \\
    69 & J000313.1-060712 & 1237672793959891223 & 18.60  & 0.278* & A2697 & 0.232  & 0.465  \\
\hline
\caption{When a spectroscopic redshift is not available, the SDSS photometric redshift is adopted instead, which is labeled with an asterisk (*).}
\label{tab:1}
\end{longtable*}

\begin{figure}
\includegraphics[width=0.45\textwidth]{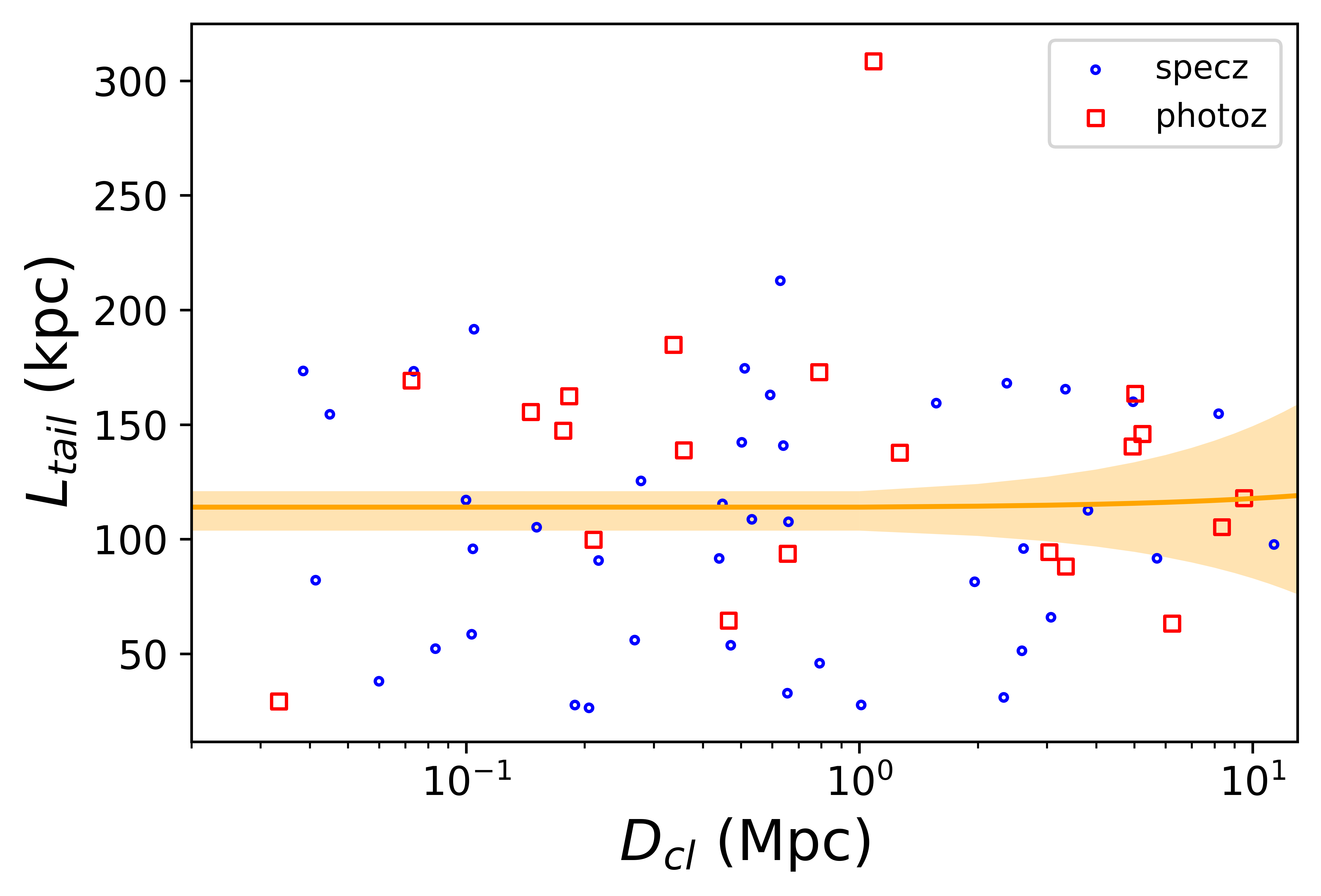}\\
\caption{The scatter of projected distances to the cluster center \textit{D}$_{cl}$ and lengths of tails. Blue dots refer to the OHTs with spectroscopic redshifts, while the red squares refer to OHTs with photometric redshifts. The orange solid line is the linear fitting result of 63 sources with cluster associations $L_{tail}/{\rm kpc} = (0.42\pm 2.73) D_{cl}/{\rm Mpc} + (113.53 \pm 8.61 )$. The shadow represents its 1$\sigma$ error range. }
\label{httail1} 
\end{figure}

To explore possible factors affecting the appearance of OHTs, we estimate the projected length of tails as $L_{tail} = (d_{ot} + fMaj_t)D_{a}$, where $D_{a}$ is the angular diameter distance. We check the relationship between the tail length and the projected distance to the cluster center in Fig.\ref{httail1}. The linear fitting result of 63 sources with cluster associations (in orange) is $L_{tail}/{\rm kpc} = (0.42\pm 2.73) D_{cl}/{\rm Mpc} + (113.53 \pm 8.61 )$. The shadow represents its 1$\sigma$ error range. We find no correlation between them. 

\begin{figure}
\includegraphics[width=0.45\textwidth]{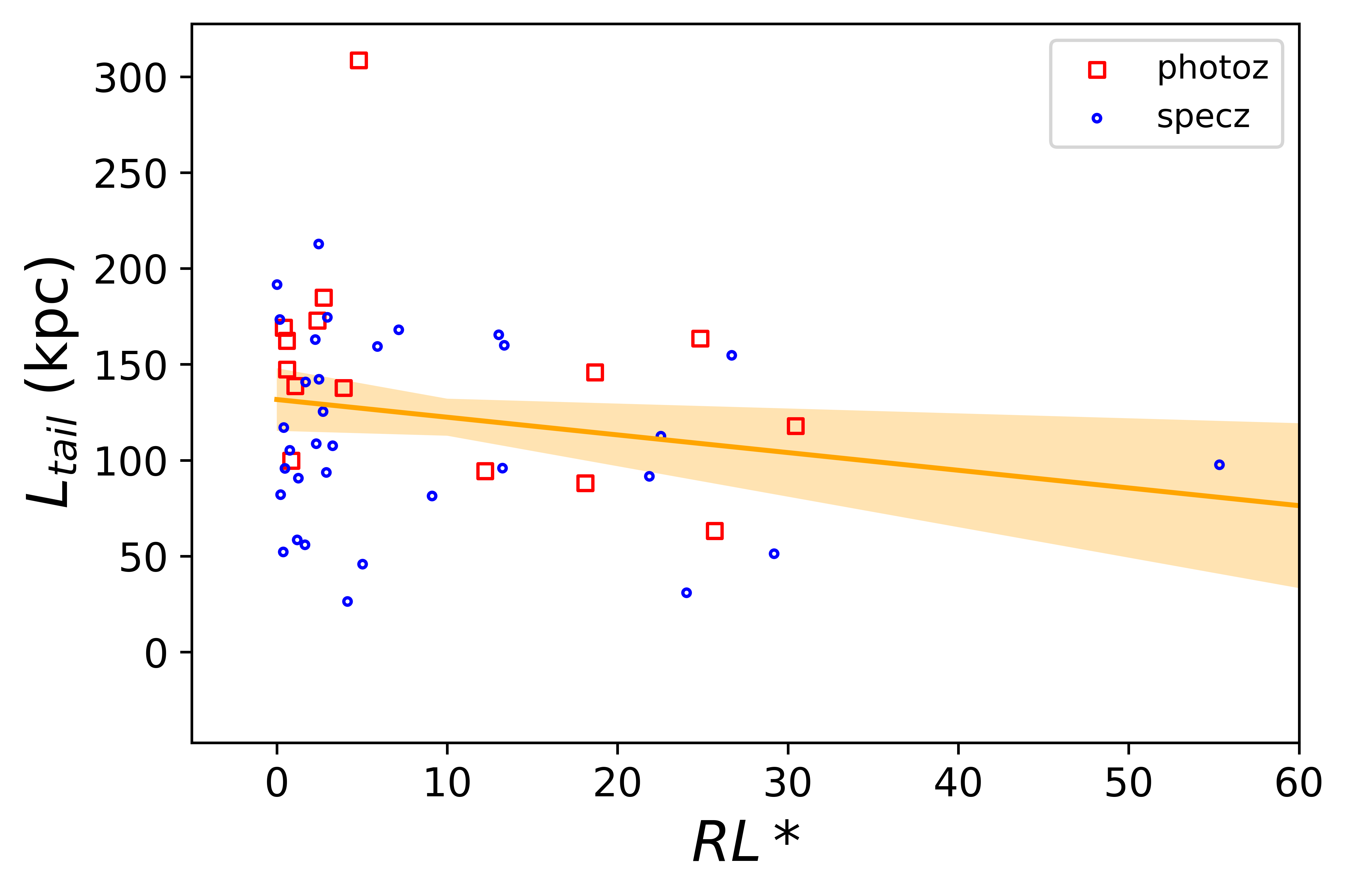}\\
\caption{The scatter of the length of tail and richness of clusters using the WHL catalog. Blue dots refer to the OHTs with spectroscopic observation, while the red squares refer to OHTs with photometric redshifts.  The orange solid line is the linear fitting result with 48 sources with WHL cluster richness, which is $L_{tail}/{\rm kpc} =  (-0.92\pm 0.67)RL* + (131.68 \pm 9.66 )$. The shadow represents its 1$\sigma$ error range.}
\label{httail3}
\end{figure}

With the richness of clusters (RL*) from the WHL2015 cluster catalog,
the tail lengths does not show a clear correlation either (Fig.\ref{httail3}).  The orange solid line is the linear fitting result of 48 sources with WHL cluster richness, $L_{tail}/{\rm kpc} =  (-0.92\pm 0.67)RL* + (131.68 \pm 9.66 )$. 
The weak anticorrelation suggests that long tails might trend to appear in poor clusters. 

\begin{figure}
\includegraphics[width=0.45\textwidth]{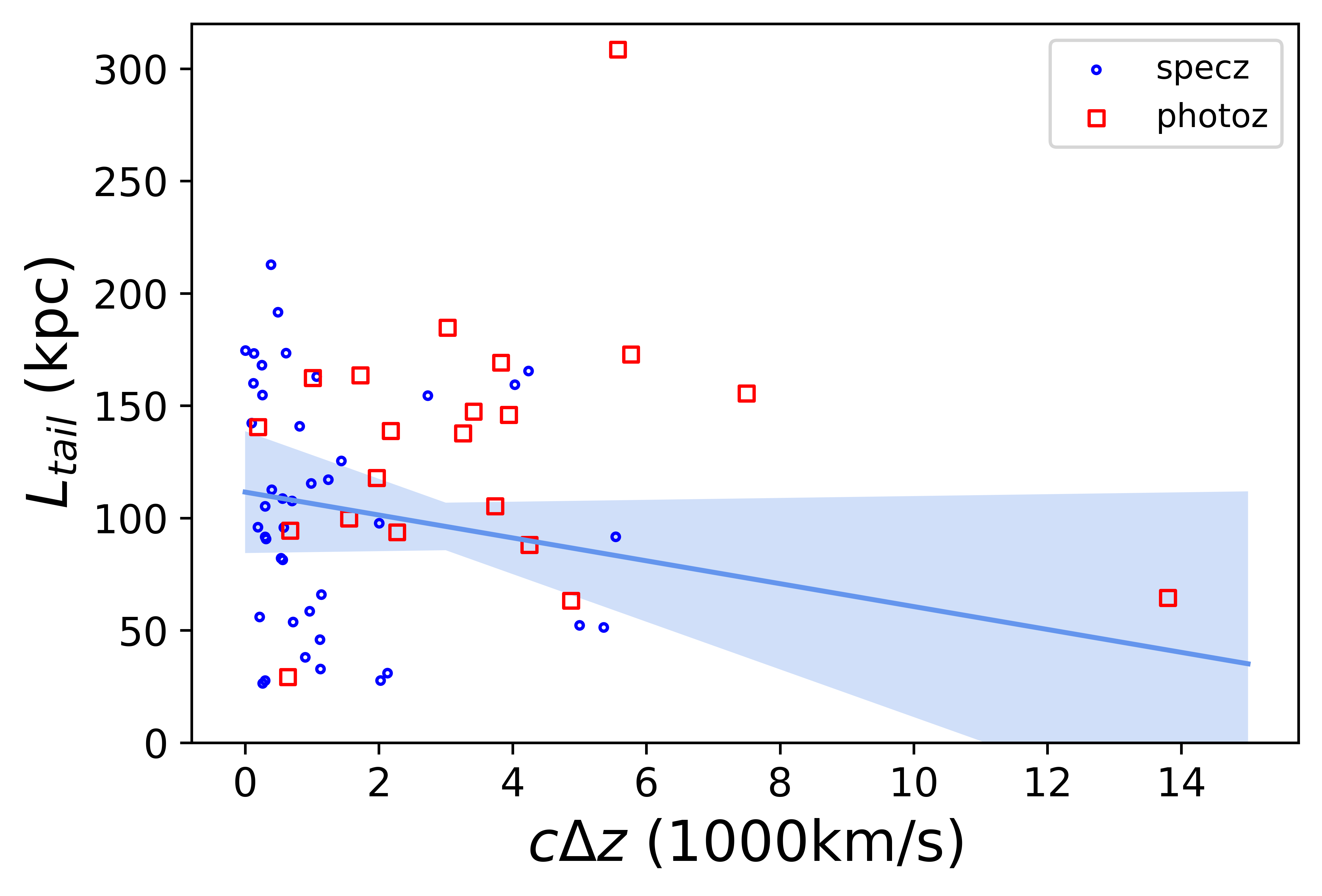}\\
\caption{The scatter of tail lengths and velocity differences c$\Delta$z. Blue dots refer to OHTs with spectroscopic redshifts, while the red squares refer to OHTs that only have photometric redshifts. The blue solid line is the linear fitting result of 48 spectroscopic sources, which is $L_{tail}/{\rm kpc} = (-5.09\pm 5.51) c\Delta z /1000{\rm km~s^{-1}}+ (111.53 \pm 10.58 )$. The shadow represents its 1$\sigma$ error range.}
\label{httail2}
\end{figure}

We also check the relationship between tail lengths and velocity differences between clusters and host galaxies c$\Delta$z as Fig.\ref{httail2} shows.  Considering that precision of redshifts is crucial in this relationship, the linear fitting is only applied to 48 OHTs with spectroscopic redshifts. The result is $L_{tail}/{\rm kpc} = (-5.09\pm 5.51) c\Delta z /1000{\rm km~s^{-1}}+ (111.53 \pm 10.58 )$. The weak anticorrelation we found is reasonable. Because the tail lengths of OHTs closely relate to the peculiar velocity of their host galaxies, the projected tail lengths observed are mainly dominated by the two velocity components projected on the sky plane. The redshift differences between galaxies and their host cluster are the radial velocity component. The larger it is, the less the other two are.

\section{Discussion and Conclusions} 
We set up an automatic procedure to search for one-side
head–tail structures in the FIRST survey catalog. It could
identify OHT sources effectively, but there are still some
sources, like NATs, WATs, one-sided jets, etc. showing a
similar pattern. It is challenging to distinguish them with a
radio image alone. After cross-checking with the SDSS
photometric catalog and galaxy cluster catalogs, we compile
an OHT catalog with 69 sources. Most of them have not been
noticed before. As the first OHT catalog, our sample provides a
fair data set for future deep-radio and optical observations.
Details of these comprehensive OHT sources will be helpful for
understanding their occurrence and properties. This catalog
could also be taken as a training sample of machine-learning
applications to find more examples in a future radio survey.

\begin{figure}
\includegraphics[width=0.45\textwidth]{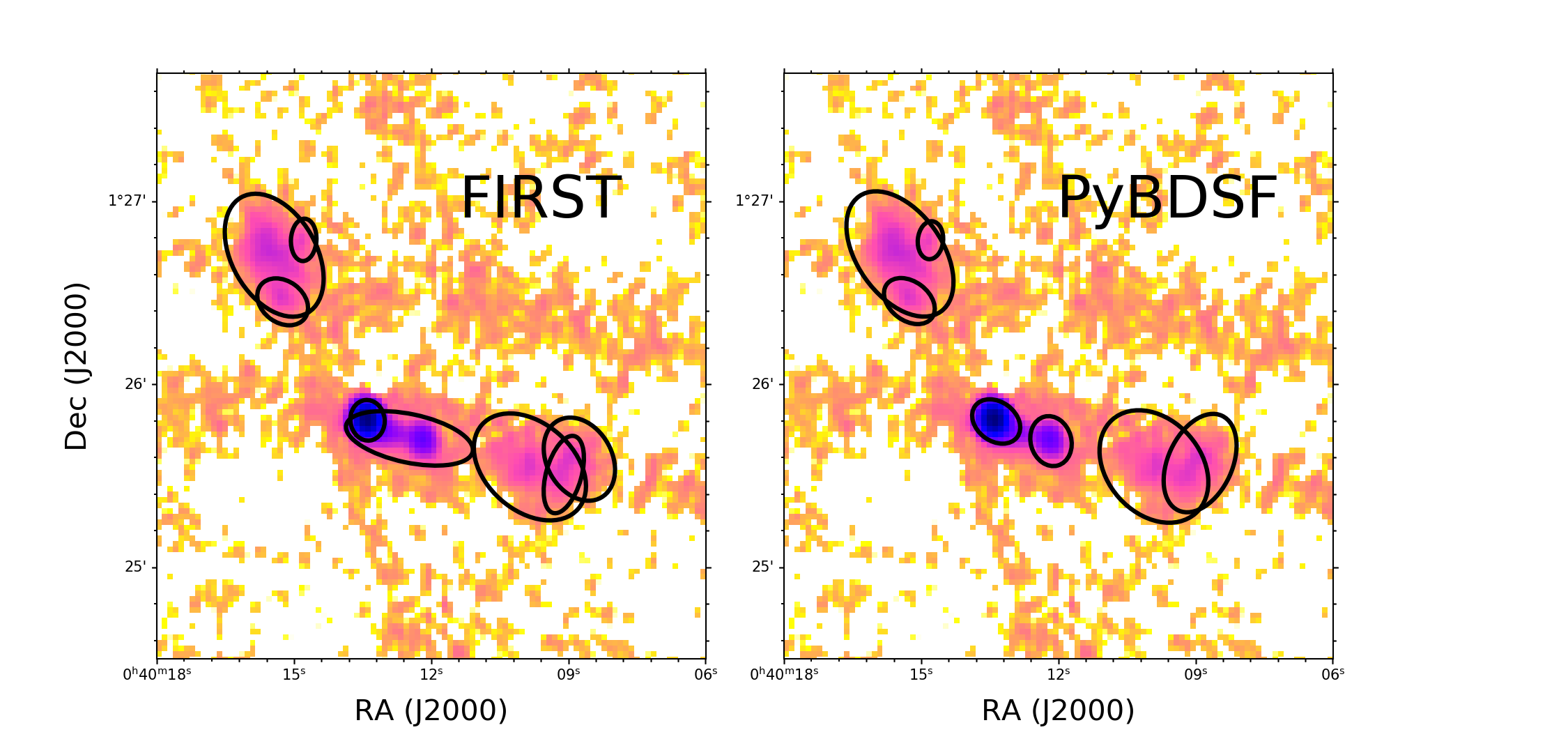}\\
\caption{The comparison of Gaussian components in FIRST catalog (left panel) and fitting by PyBDSF (right panel) in the same field of J004013.5+012546, which is one of the 115 OHT candidates but is dropped by the visual check. Each panel is in the size 3$^\prime$ $\times$ 3$^\prime$.}
\label{compa}
\end{figure}

We also notice that Gaussian components in the FIRST catalog do not always provide a good morphological description of radio sources. With a modern fitting procedure like PyBDSF\citep{2015ascl.soft02007M}, the source list could describe extended emission better, and the performance of our procedure could then be further improved. A example case, J004013.5+012546, is given by Fig.\ref{compa}. It is one of the 115 OHT candidates but is recognized as a jet lobe by the visual check, then dropped. Its central part shows an HT-like structure in the FIRST fitting result (left panel), while the PyBDSF fitting regions (right panel) give a more reasonable estimate. However, updating the whole FIRST catalog with the new fitting procedure is beyond the goal of this paper. 

With this sample, we confirm that most OHTs are in the gravitational potential wells of clusters. 
The lengths of their tails do not correlate with the projection distance to the center of the nearest galaxy clusters.
But they show weak anticorrelation with the cluster richness, and are inversely proportional to the radial velocity differences between clusters and host galaxies.
There are still some sources that are in the
outskirts of clusters or even fully outside known clusters.
Further radio observations with high angular resolution and
spectral maps are necessary to reveal more details. 

Upcoming radio surveys, like the LoTSS \citep{2017A&A...598A.104S}, the Australian Square Kilometre Array Pathfinder survey\citep{2007PASA...24..174J}, the Jansky VLA Sky Survey \citep{2020PASP..132c5001L}, and the Square Kilometer Array (https://www.skatelescope.org/) will discover more HT sources and provide better chances to study and understand this special type of radio galaxies.

\section*{Acknowledgments}
We sincerely thank the anonymous referee for the
constructive comments. We also thank Hengrui Ma for his
contribution at the early stage of this project, and are grateful
to Pietro Reviglio for his generous help and valuable
discussions. This work was supported by the Bureau of
International Cooperation, Chinese Academy of Sciences,
under the grant GJHZ1864. R.J.v.W. acknowledges
support from the CAS-NWO program for radio
astronomy with project number 629.001.024, which is
financed by the Netherlands Organisation for Scientific
Research (NWO).

\appendix
\section{identification charts of OHT candidates}
To show the morphology of OHT candidates, we list
identification charts of all 69 sources in our sample. The radio contours overlap on top of SDSS \textit{r}-band images. The ellipses on the map represent source shapes from the
FIRST catalog. The recognized heads and tails are labeled with
a small “h” and “t” correspondingly. With identification charts,
we can compare radio morphologies and FIRST fitting results
directly, and check if assigned optical counterparts are chance
alignments. With these images, we confirm that our procedure
of OHT identification is generally consistent with the visual
check.

\begin{longtable}{cccc}
\includegraphics[width=0.24\textwidth]{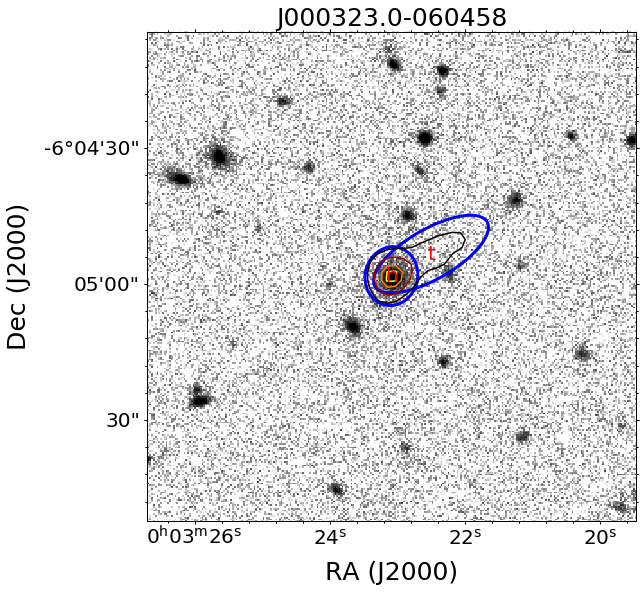}  &
\includegraphics[width=0.24\textwidth]{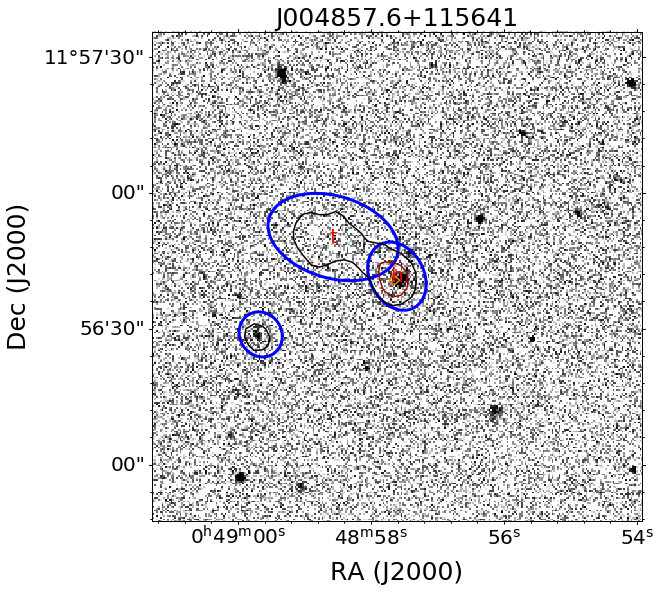}  &
\includegraphics[width=0.24\textwidth]{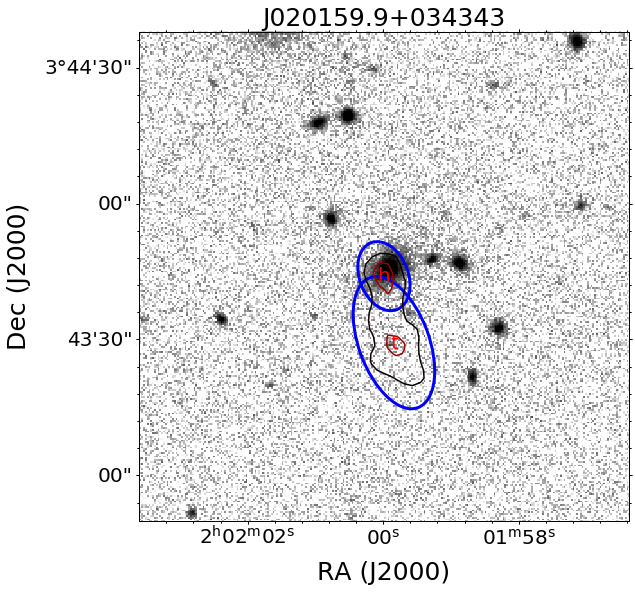}  &
\includegraphics[width=0.24\textwidth]{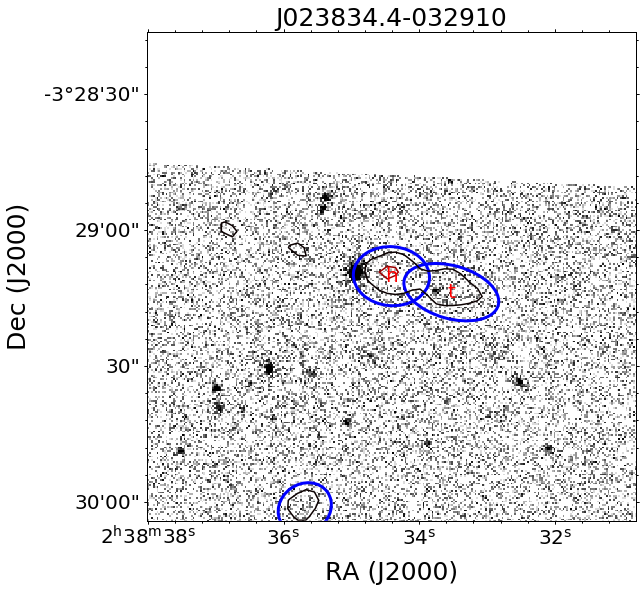} \\
1 & 2 & 3&4 \\
\includegraphics[width=0.24\textwidth]{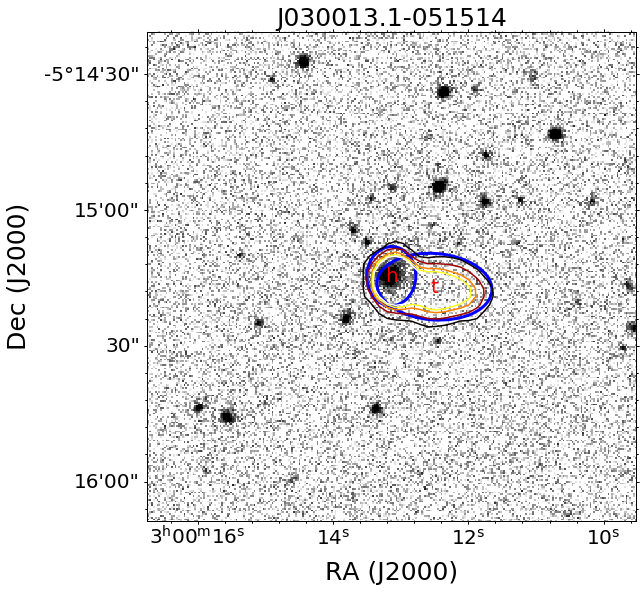} &
\includegraphics[width=0.24\textwidth]{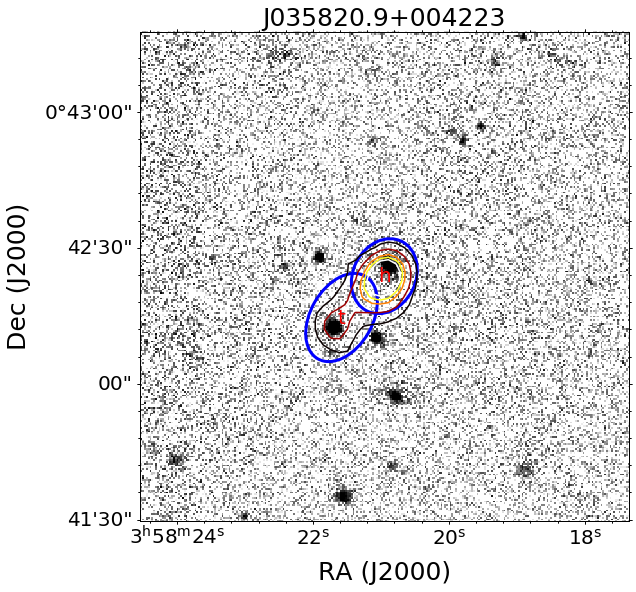} &
\includegraphics[width=0.24\textwidth]{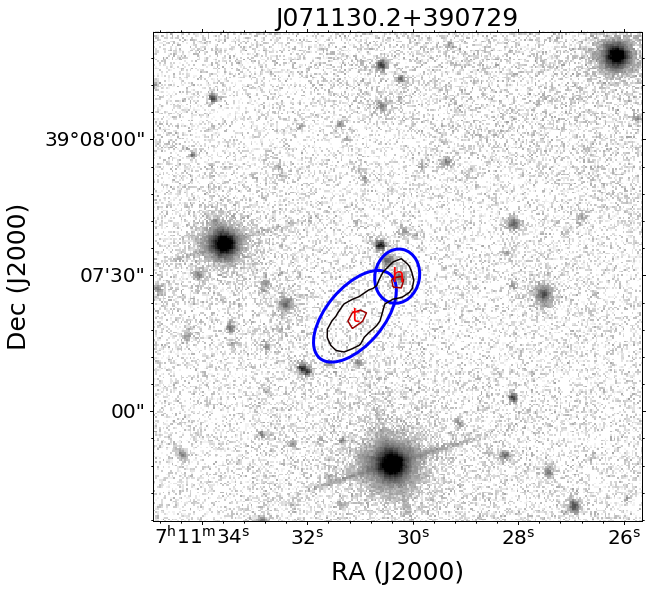} &
\includegraphics[width=0.24\textwidth]{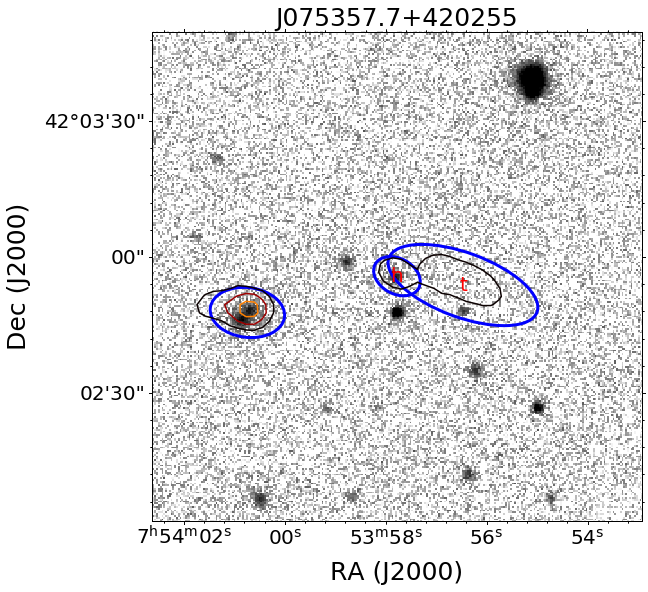} \\
5 & 6 & 7 &8\\
\includegraphics[width=0.24\textwidth]{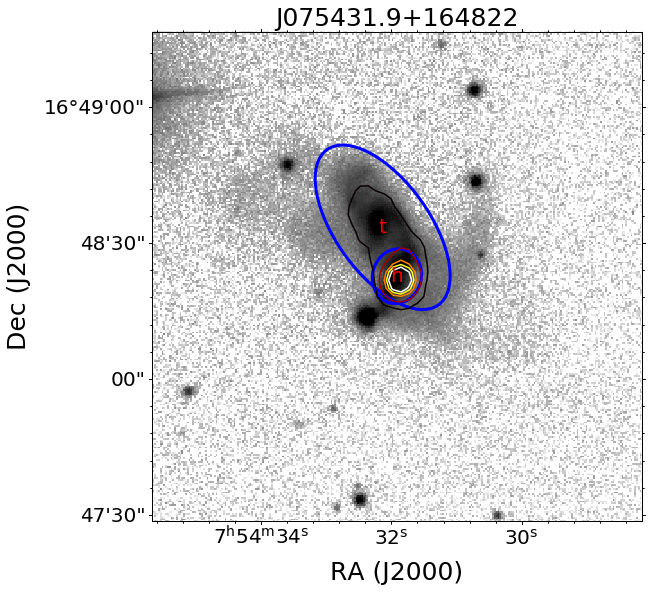} &
\includegraphics[width=0.24\textwidth]{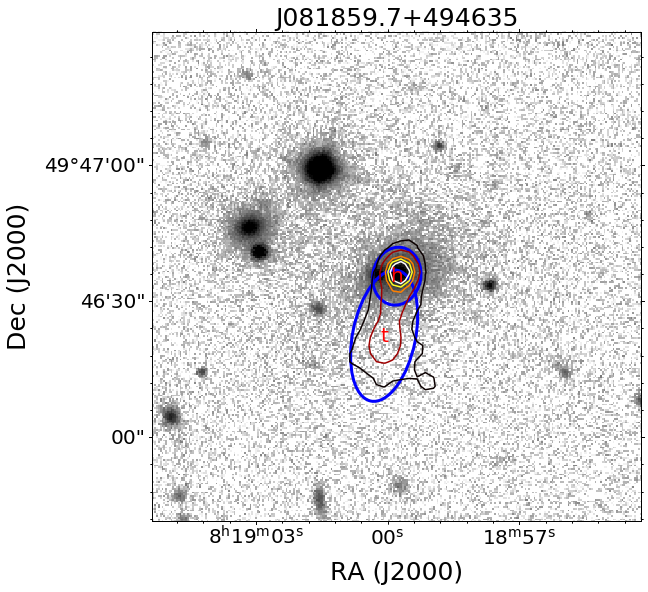} &
\includegraphics[width=0.24\textwidth]{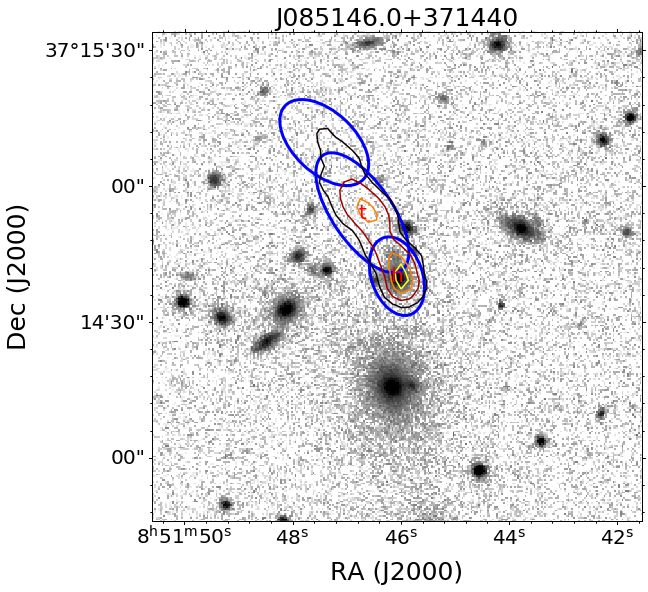} &
\includegraphics[width=0.24\textwidth]{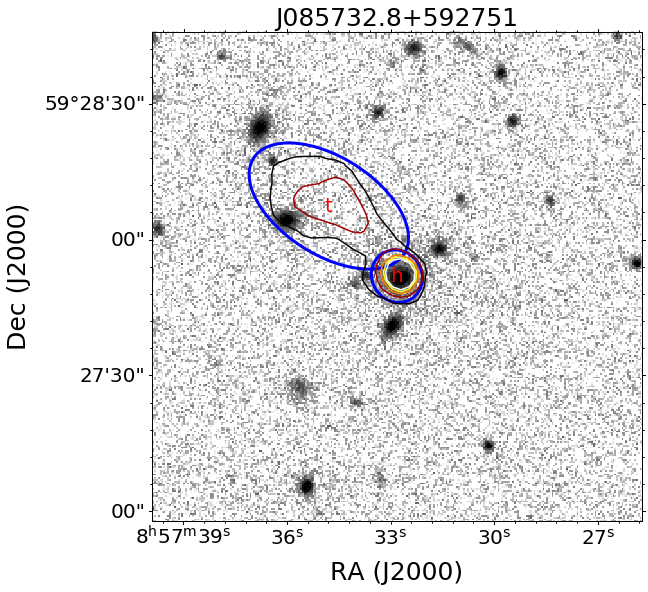} \\
9 & 10 & 11 &12\\

\includegraphics[width=0.24\textwidth]{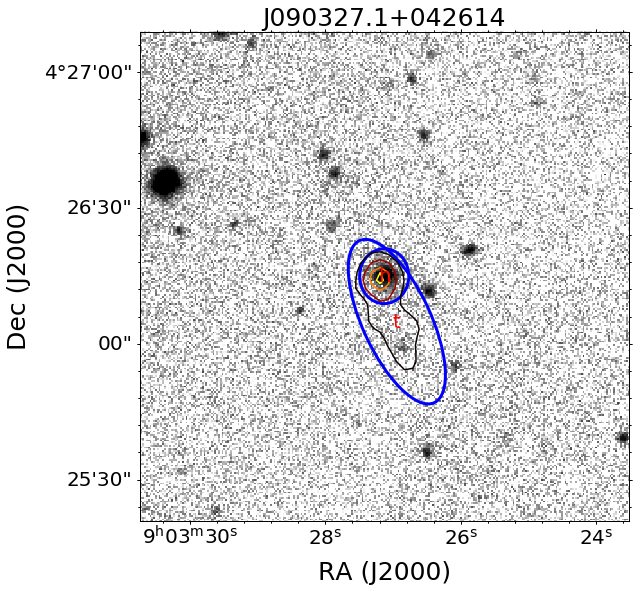} &
\includegraphics[width=0.24\textwidth]{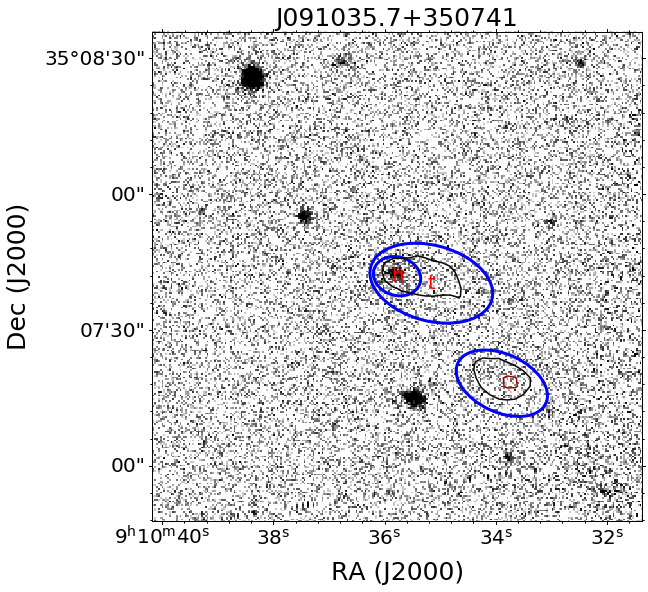} &
\includegraphics[width=0.24\textwidth]{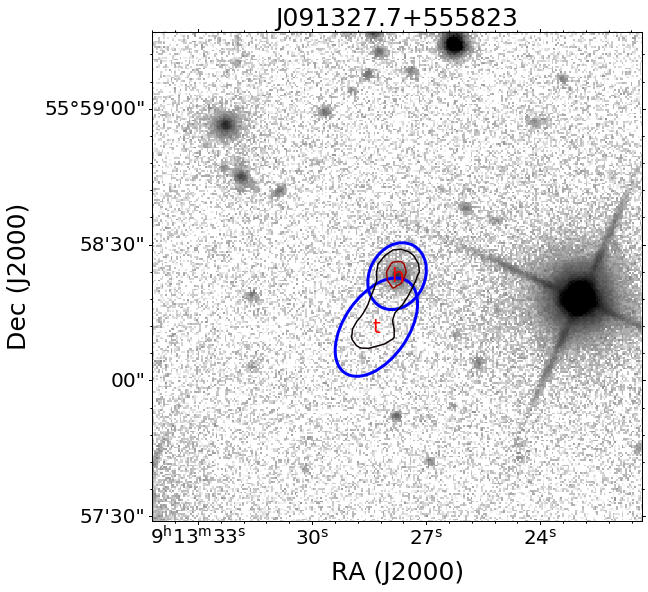} &
\includegraphics[width=0.24\textwidth]{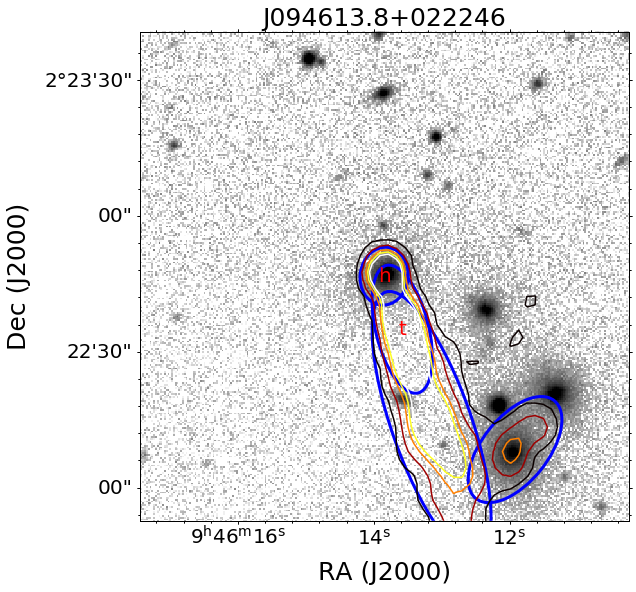} \\
13 & 14 & 15 &16\\

\includegraphics[width=0.24\textwidth]{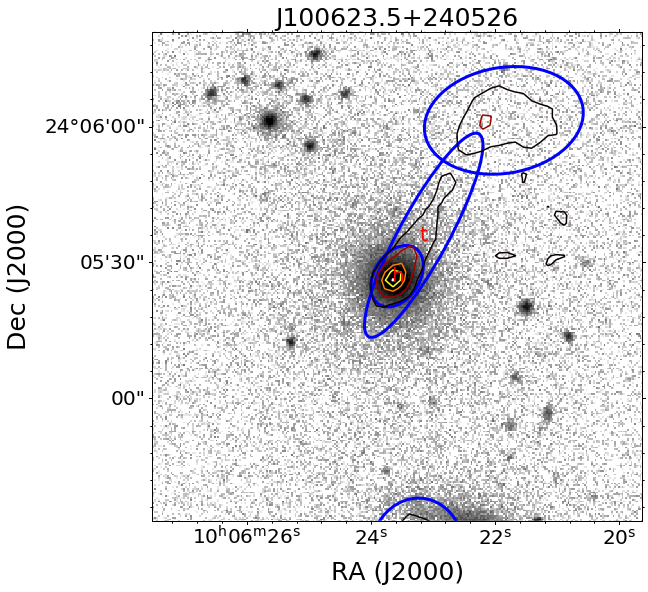} &
\includegraphics[width=0.24\textwidth]{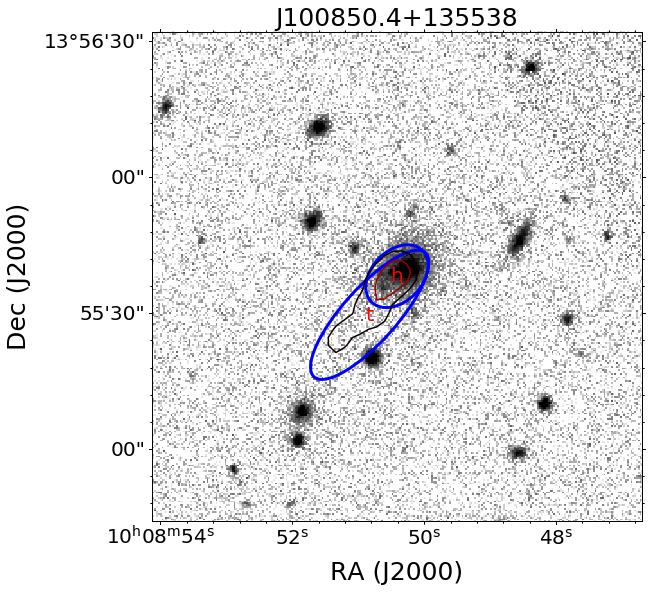} &
\includegraphics[width=0.24\textwidth]{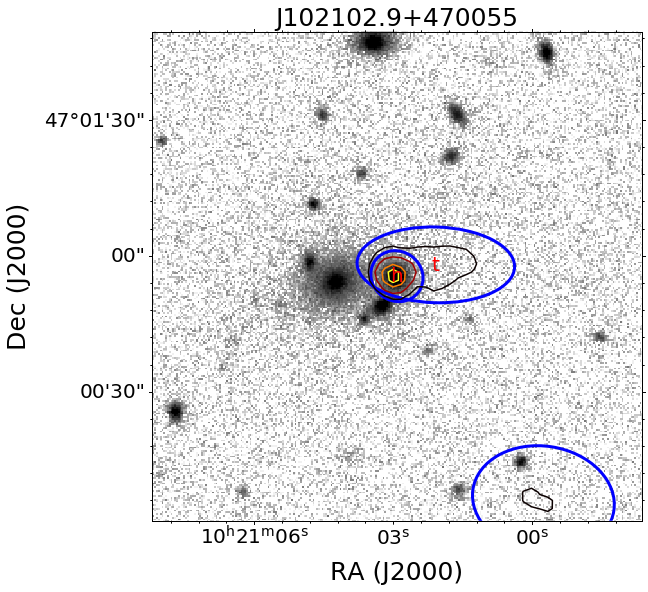} & 
\includegraphics[width=0.24\textwidth]{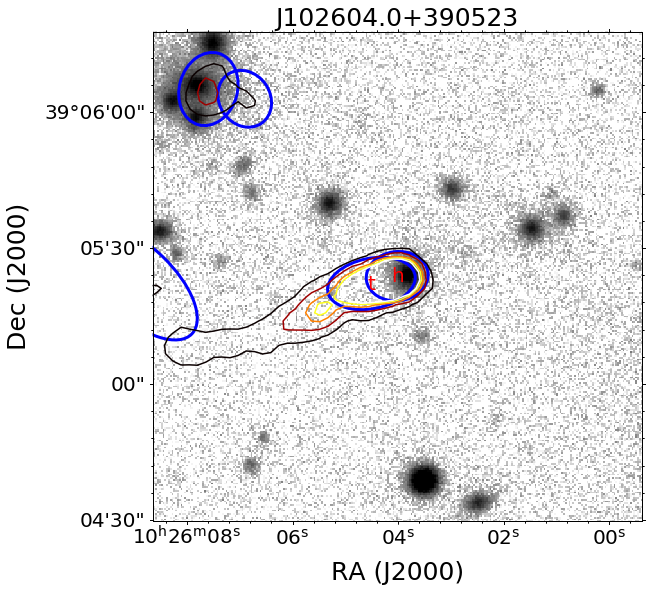} \\
17 & 18 & 19 &20\\
\includegraphics[width=0.24\textwidth]{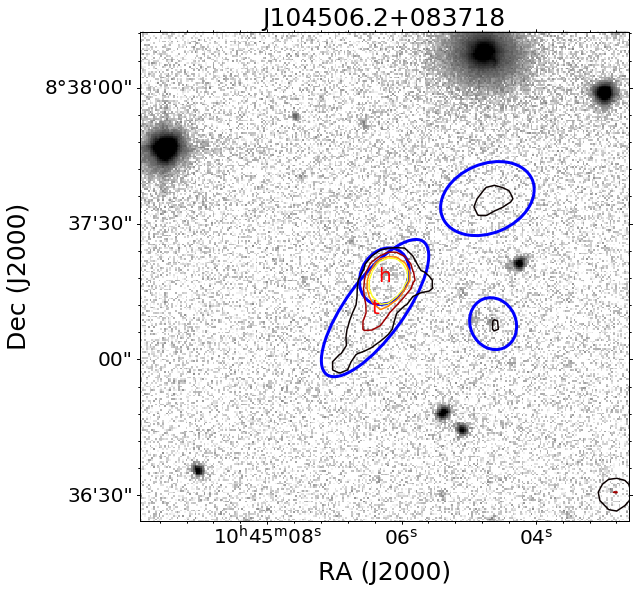} &
\includegraphics[width=0.24\textwidth]{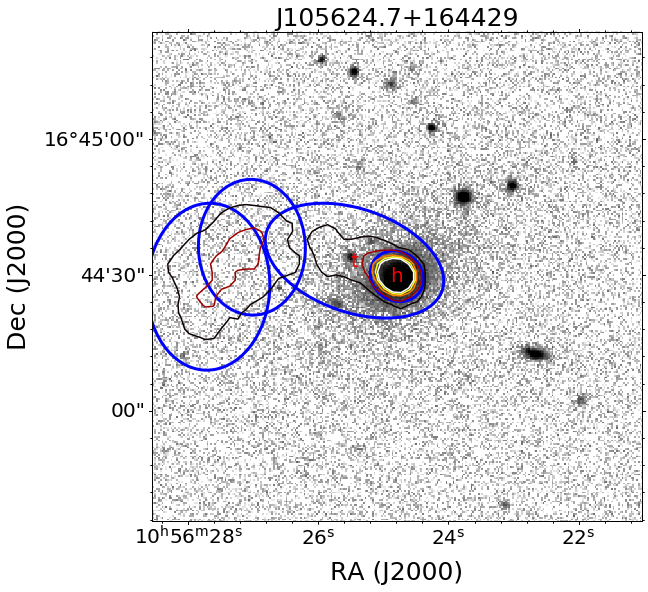} &
\includegraphics[width=0.24\textwidth]{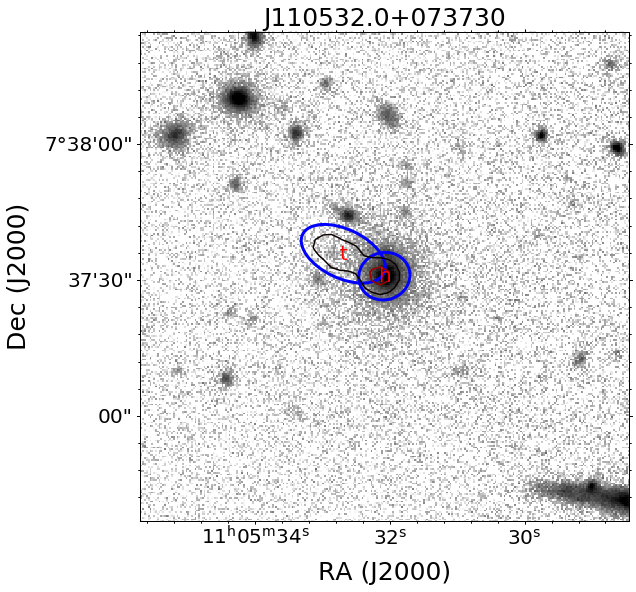} &
\includegraphics[width=0.24\textwidth]{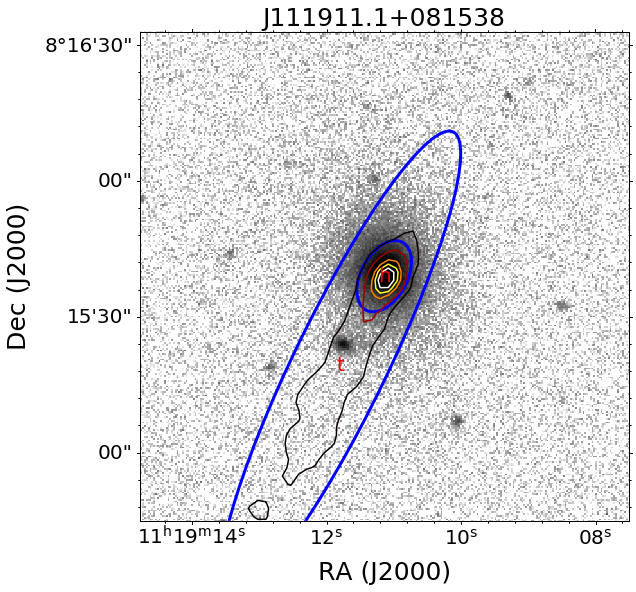} \\
21 & 22 & 23 &24\\
\includegraphics[width=0.24\textwidth]{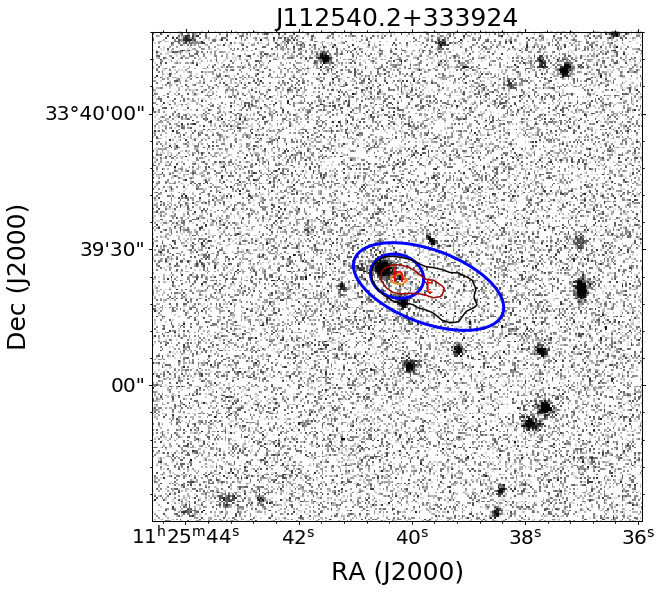} &
\includegraphics[width=0.24\textwidth]{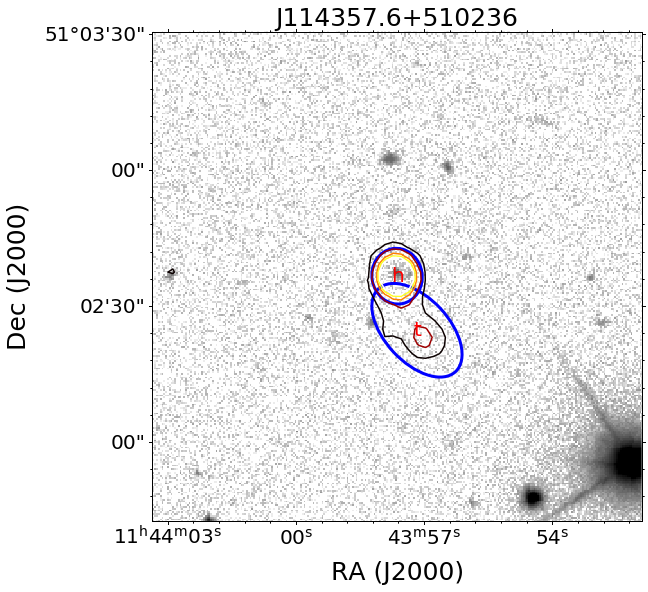} &
\includegraphics[width=0.24\textwidth]{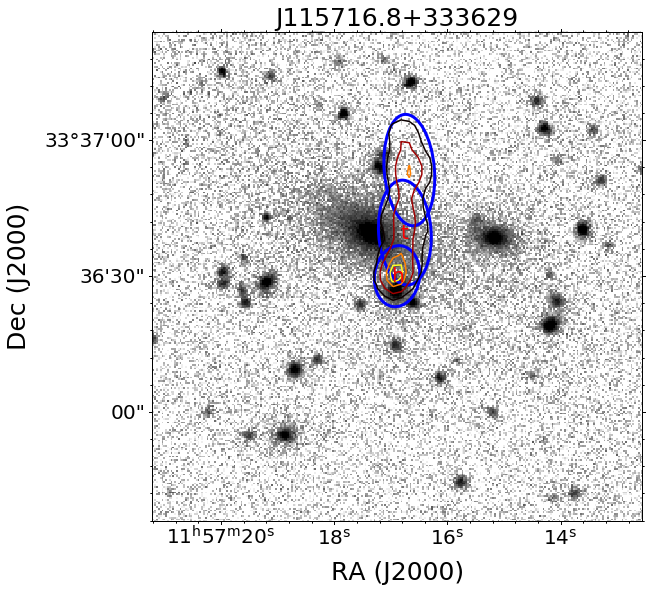} &
\includegraphics[width=0.24\textwidth]{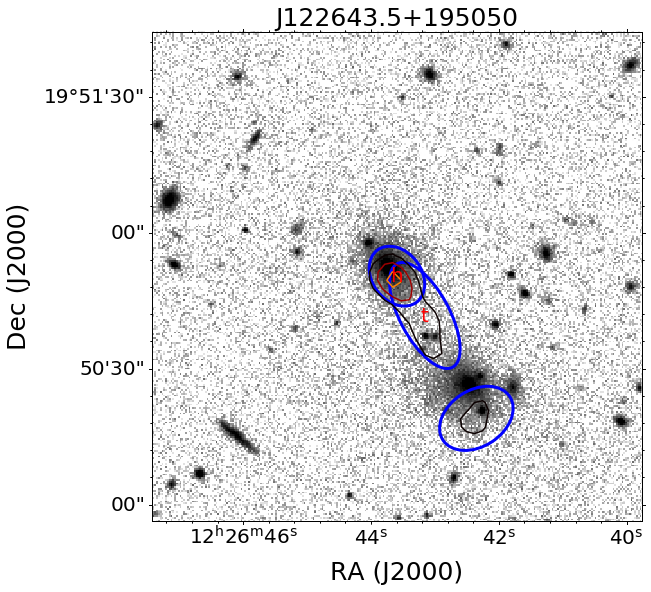} \\
25 & 26 & 27 &28\\
\includegraphics[width=0.24\textwidth]{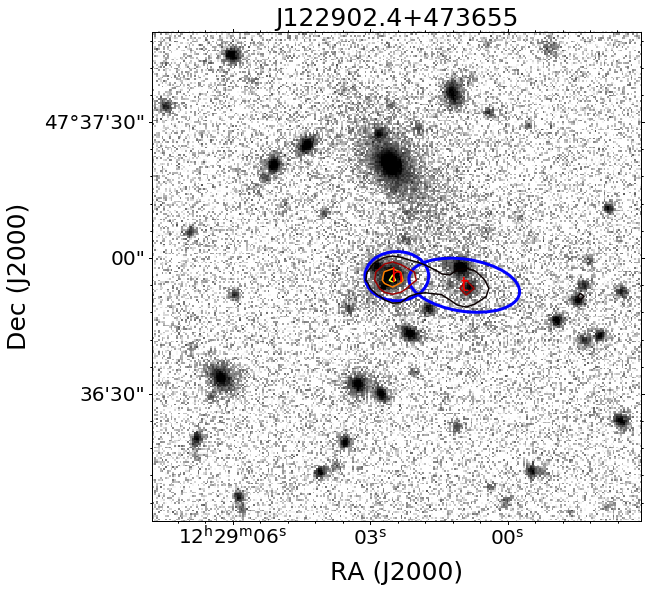} &
\includegraphics[width=0.24\textwidth]{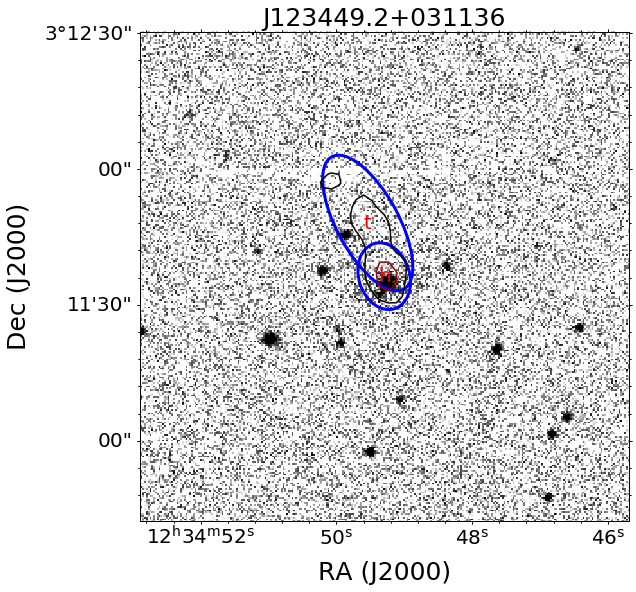} &
\includegraphics[width=0.24\textwidth]{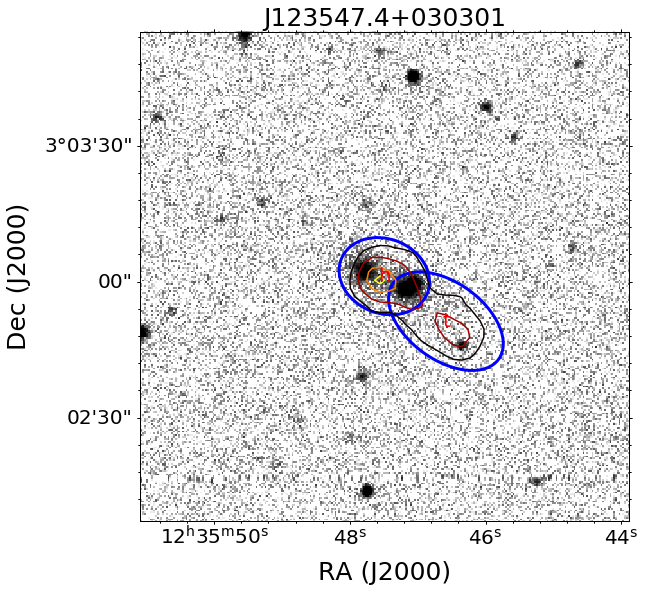} &
\includegraphics[width=0.24\textwidth]{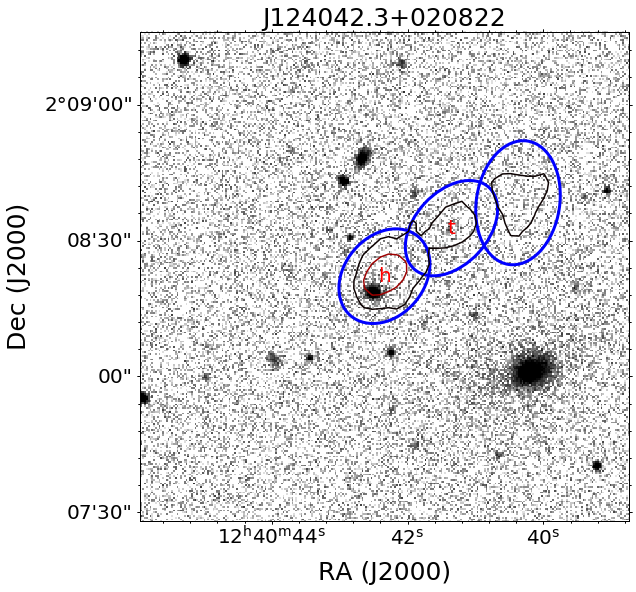} \\
29 & 30 & 31 &32\\
\includegraphics[width=0.24\textwidth]{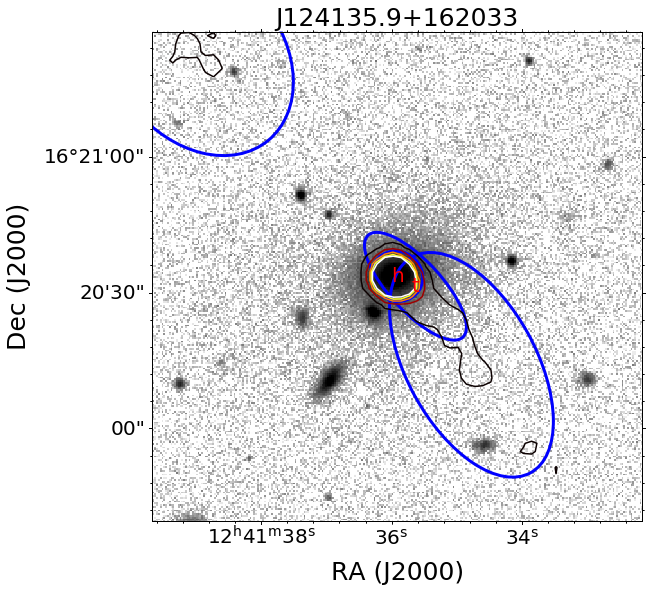} &
\includegraphics[width=0.24\textwidth]{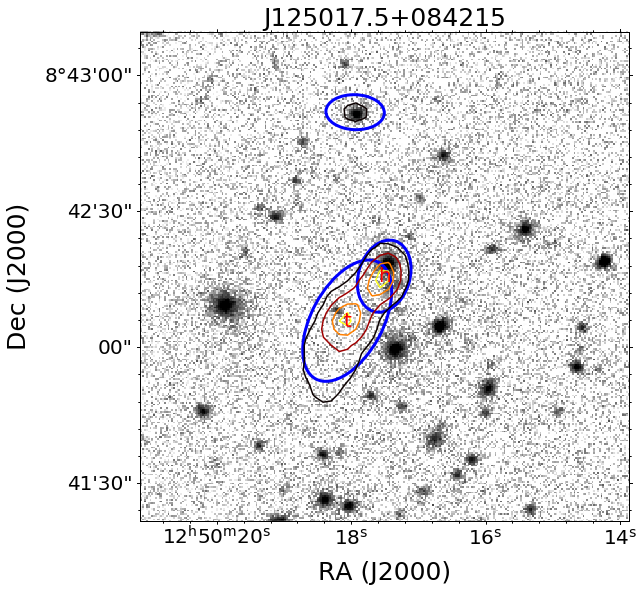} &
\includegraphics[width=0.24\textwidth]{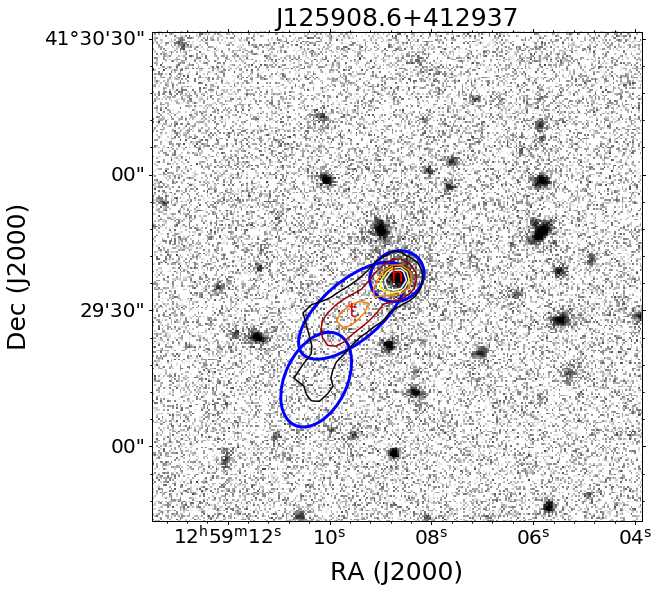} &
\includegraphics[width=0.24\textwidth]{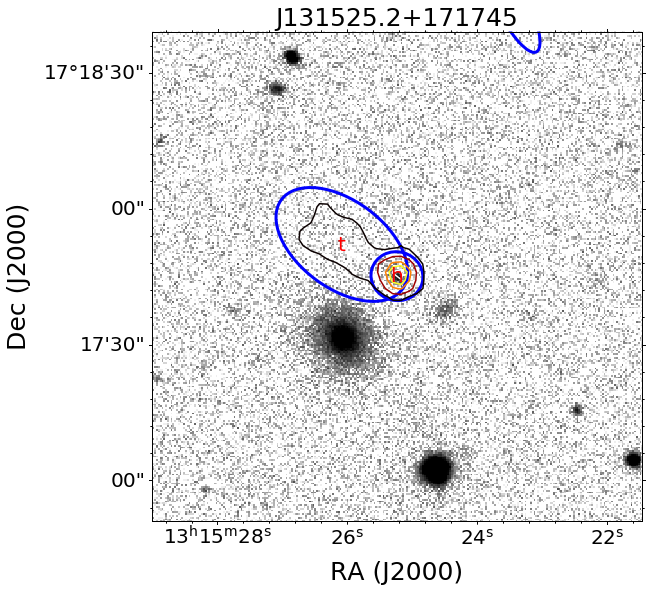} \\
33 & 34 & 35 &36\\
\includegraphics[width=0.24\textwidth]{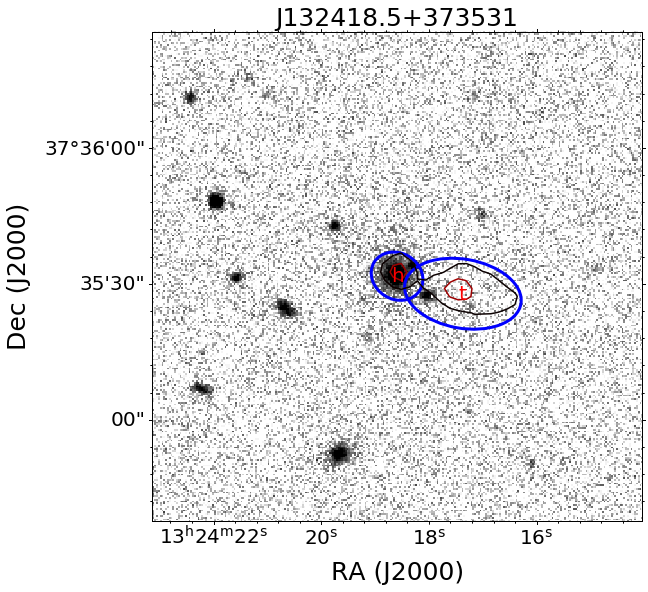} &
\includegraphics[width=0.24\textwidth]{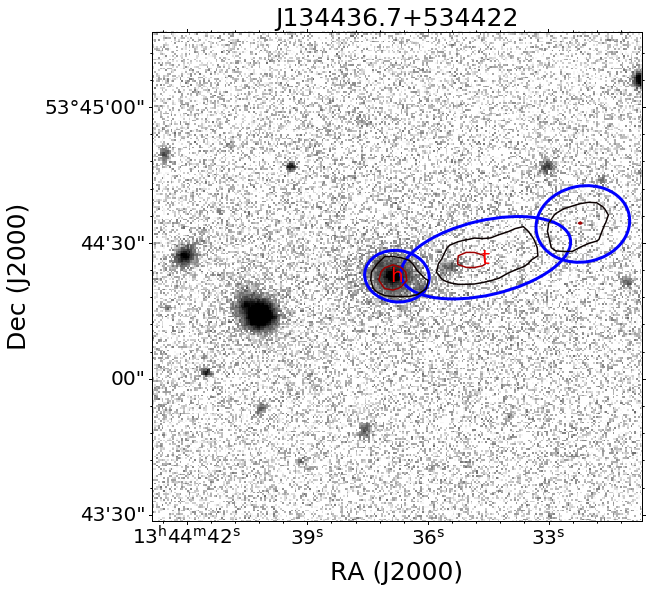} &
\includegraphics[width=0.24\textwidth]{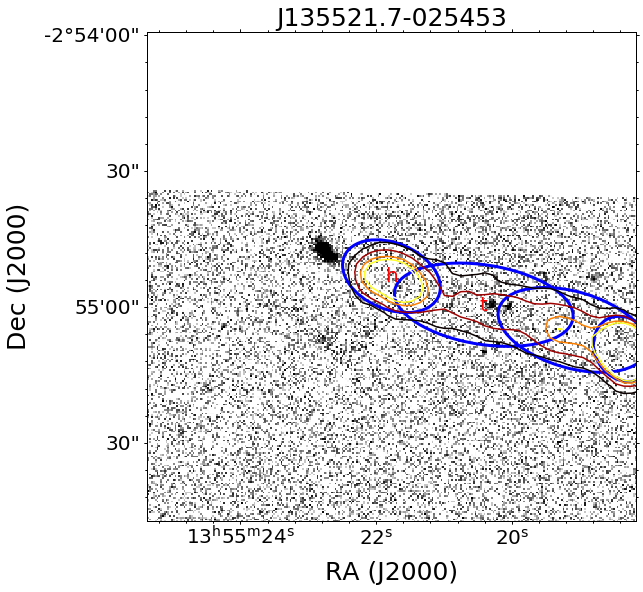} &
\includegraphics[width=0.24\textwidth]{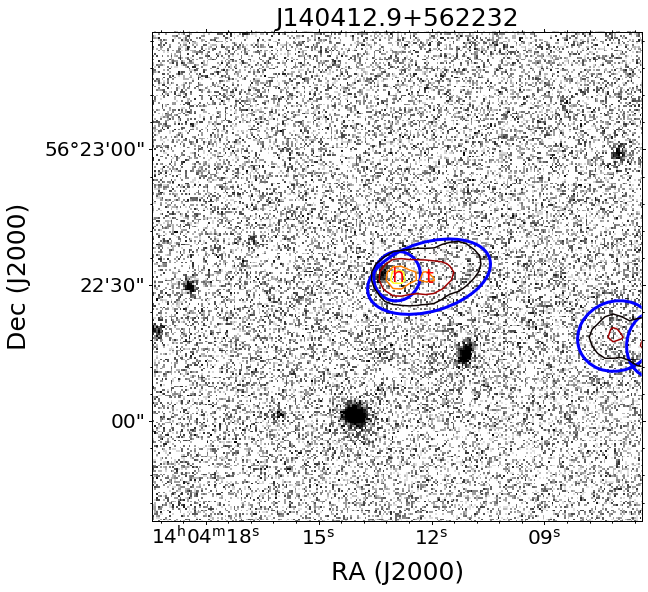} \\
37 & 38 & 39 &40\\
\includegraphics[width=0.24\textwidth]{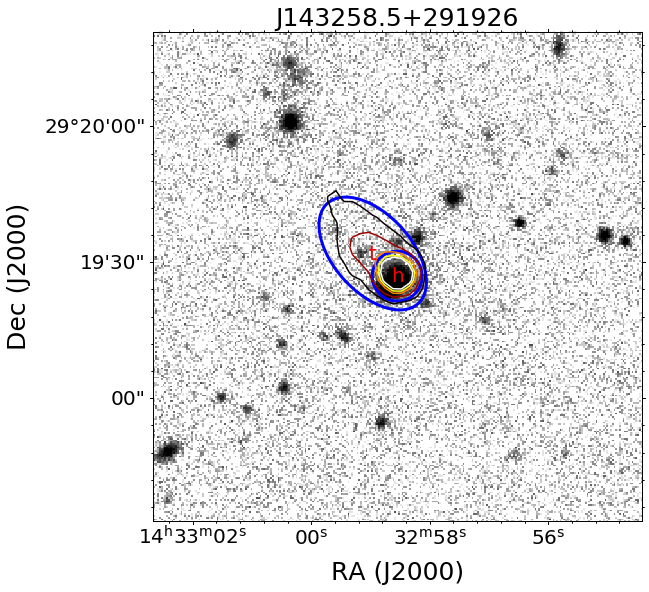} &
\includegraphics[width=0.24\textwidth]{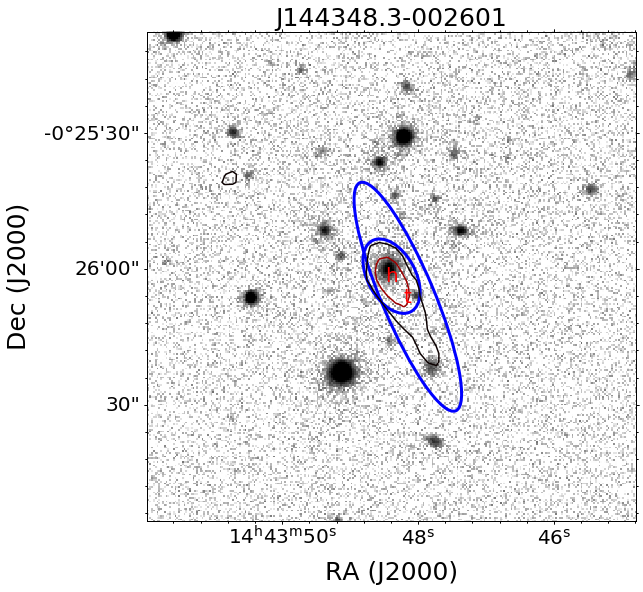} &
\includegraphics[width=0.24\textwidth]{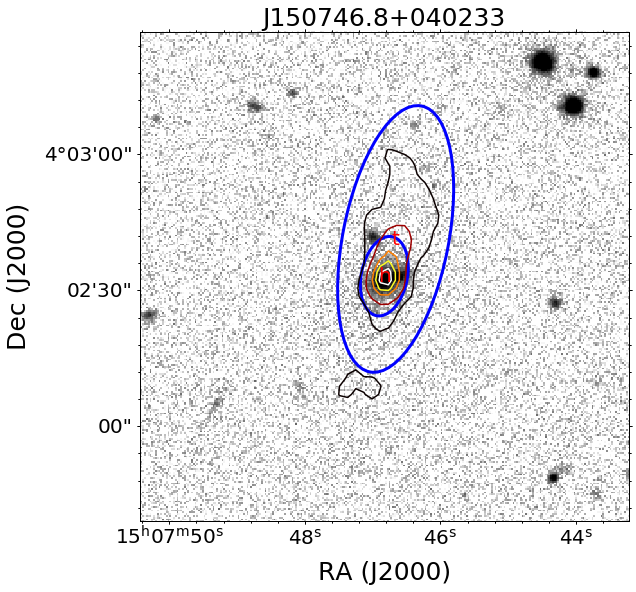} &
\includegraphics[width=0.24\textwidth]{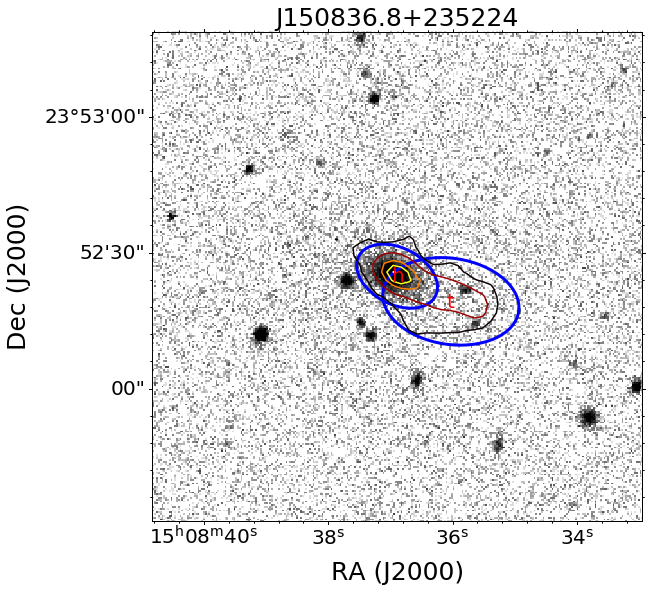} \\
41 & 42 & 43 &44\\
\includegraphics[width=0.24\textwidth]{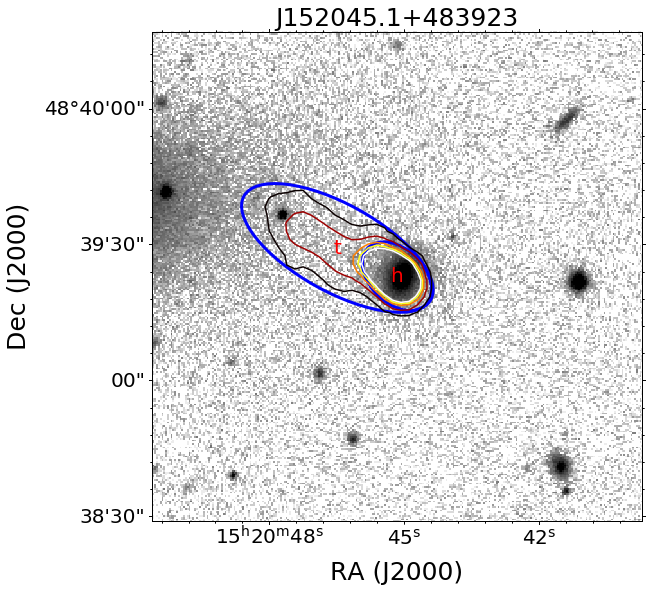} &
\includegraphics[width=0.24\textwidth]{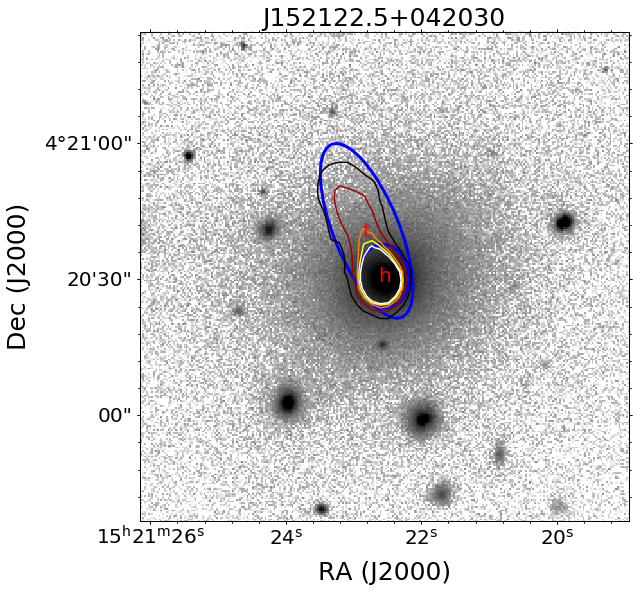} &
\includegraphics[width=0.24\textwidth]{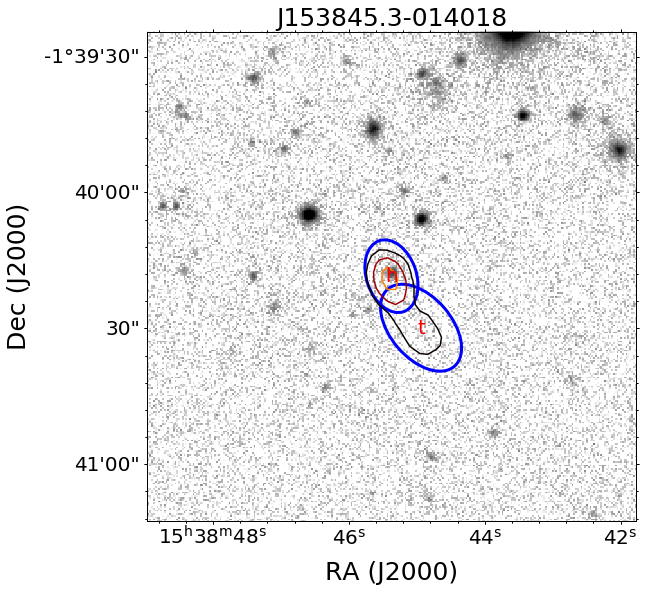} &
\includegraphics[width=0.24\textwidth]{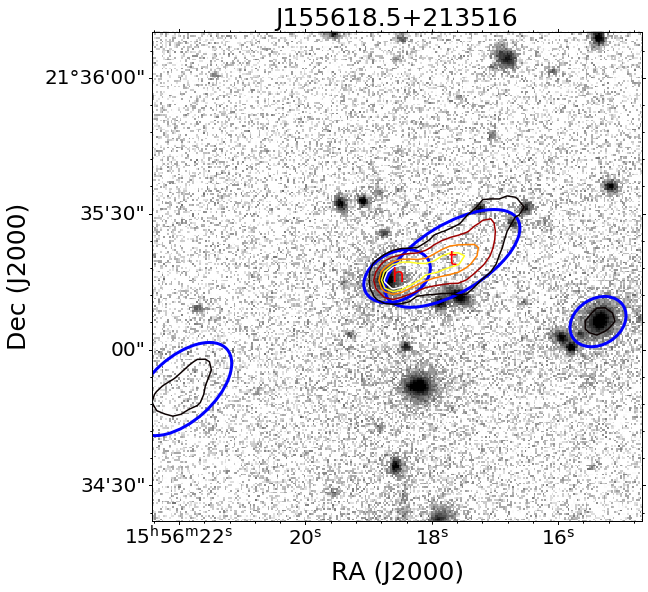} \\
45 & 46 & 47 &48\\
\includegraphics[width=0.24\textwidth]{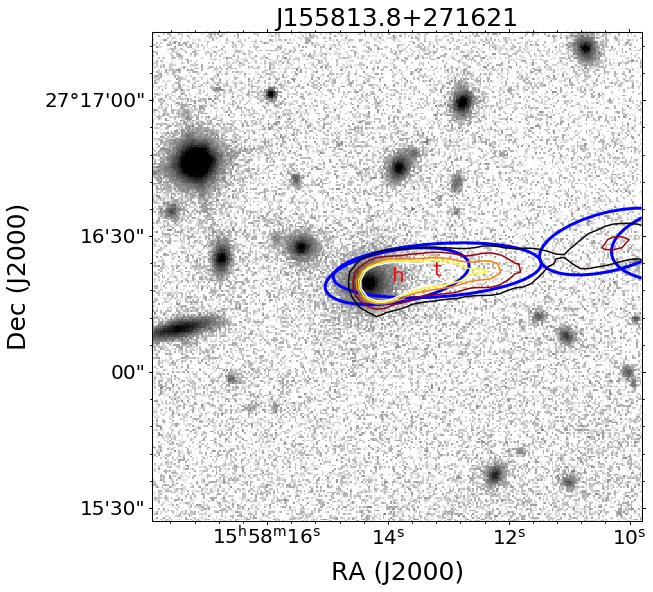} &
\includegraphics[width=0.24\textwidth]{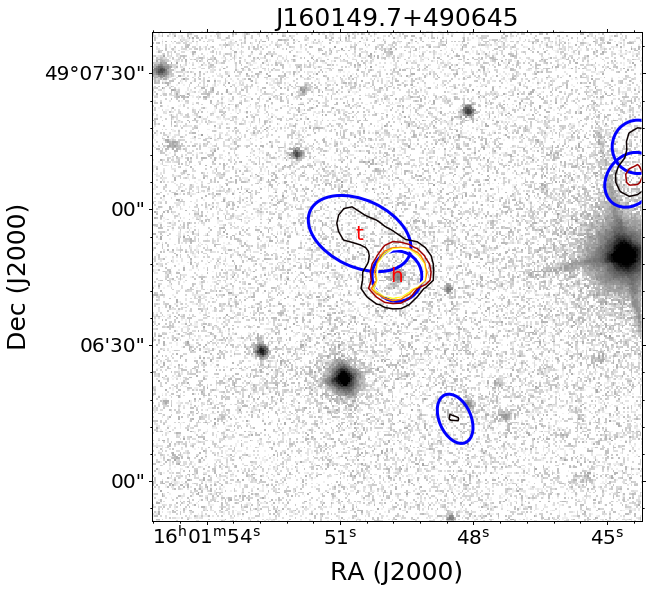} &
\includegraphics[width=0.24\textwidth]{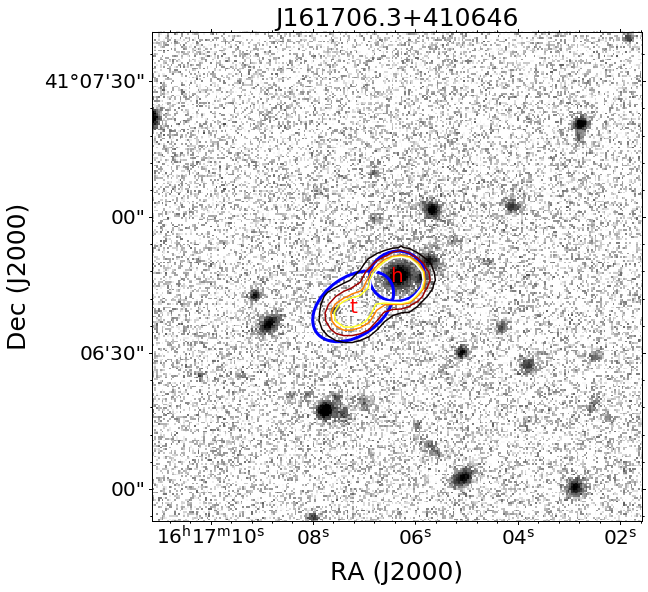} &
\includegraphics[width=0.24\textwidth]{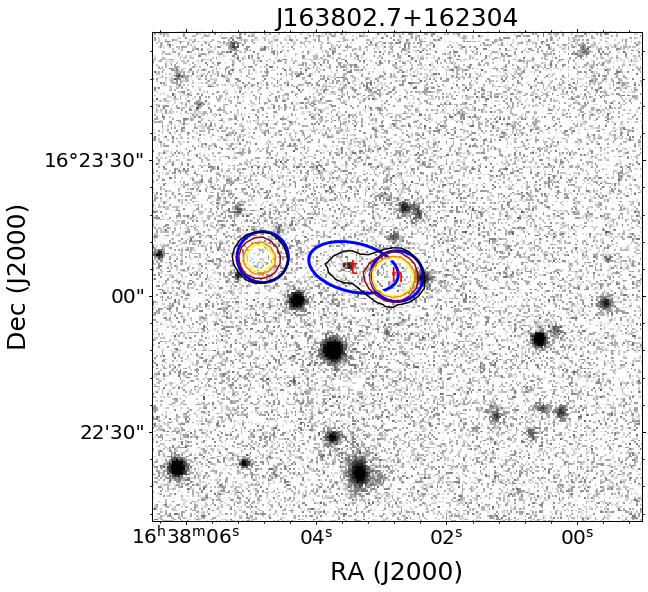} \\
49 & 50 & 51 &52\\
\includegraphics[width=0.24\textwidth]{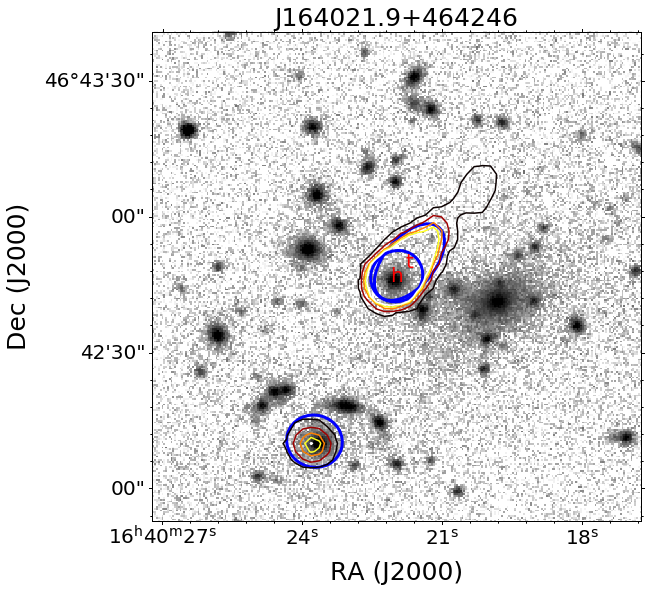} &
\includegraphics[width=0.24\textwidth]{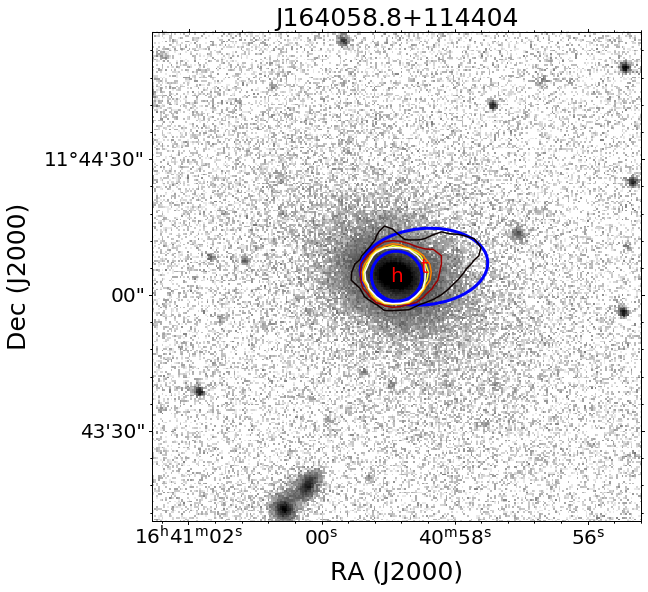} &
\includegraphics[width=0.24\textwidth]{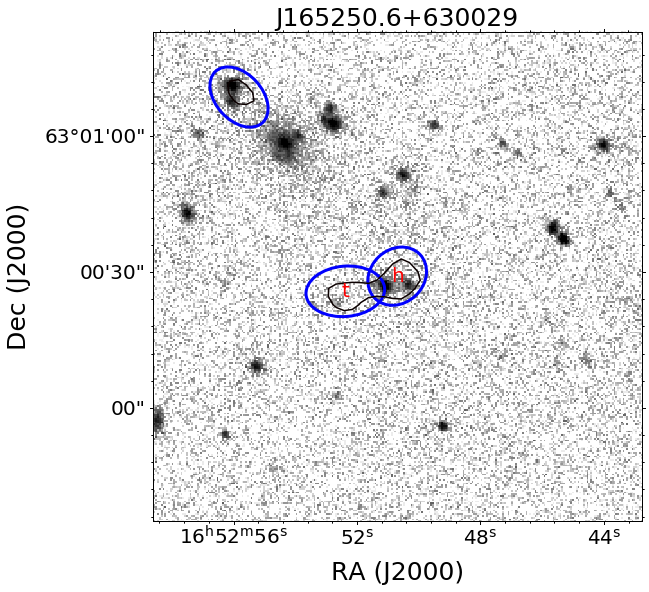} &
\includegraphics[width=0.24\textwidth]{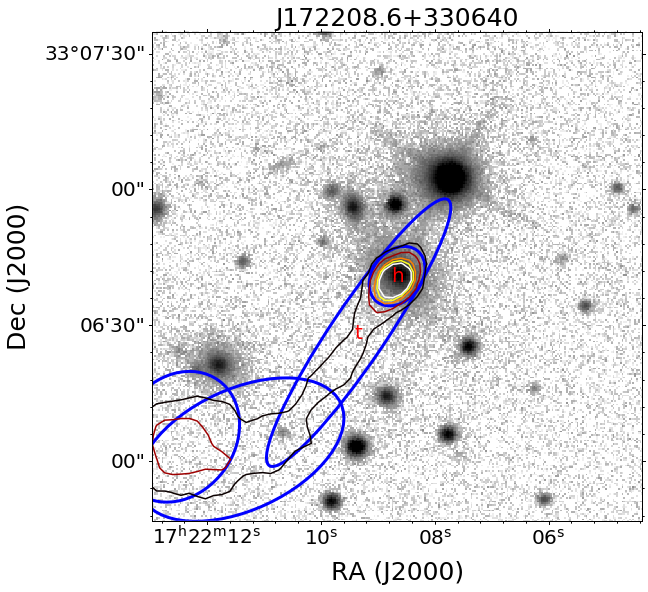} \\
53 & 54 & 55 &56\\
\includegraphics[width=0.24\textwidth]{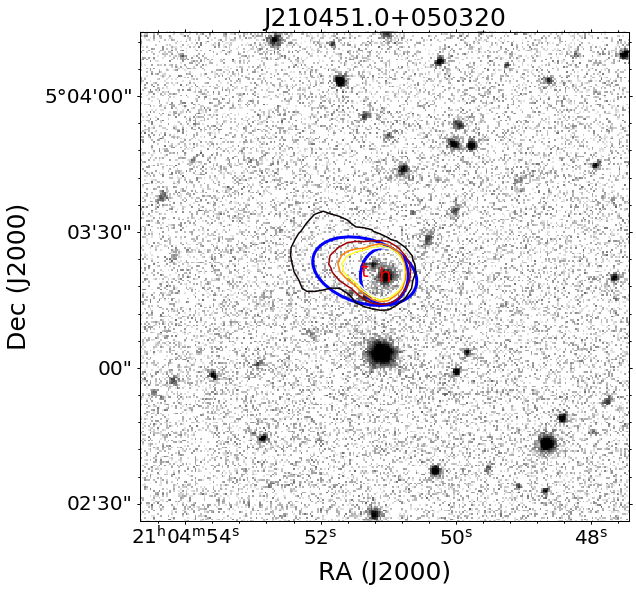} &
\includegraphics[width=0.24\textwidth]{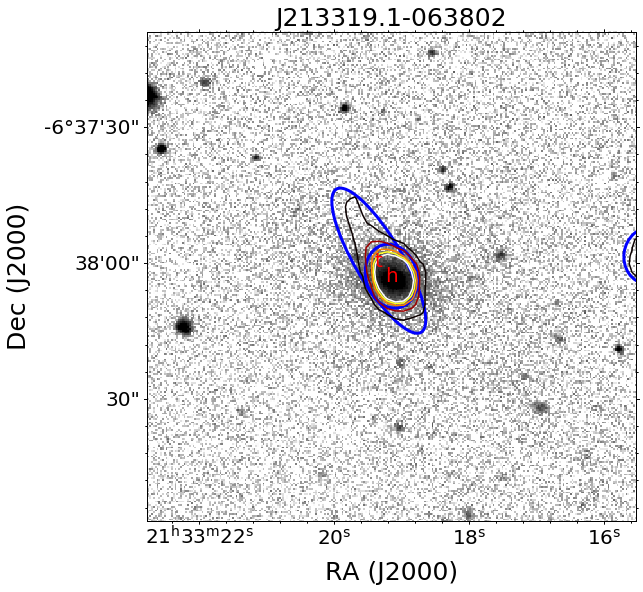} &
\includegraphics[width=0.24\textwidth]{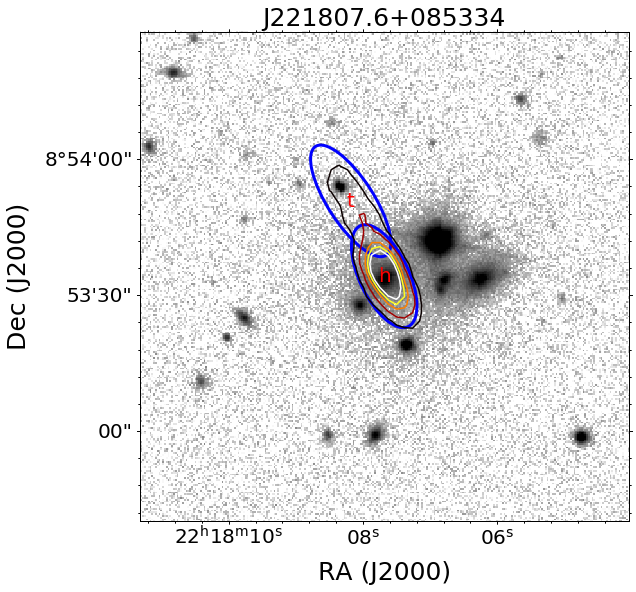} &
\includegraphics[width=0.24\textwidth]{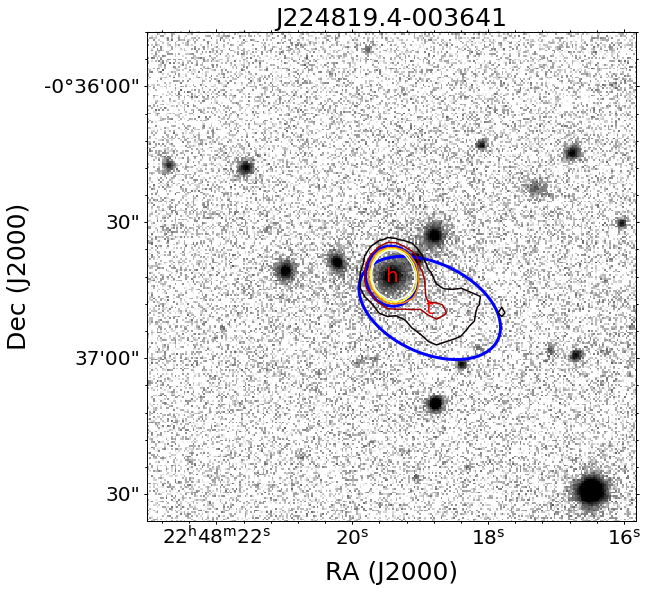} \\
57 & 58 & 59 &60\\

\includegraphics[width=0.24\textwidth]{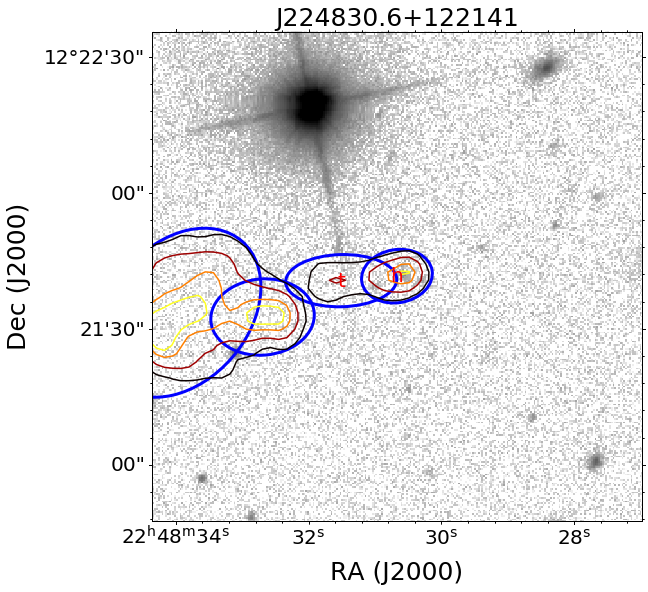} &
\includegraphics[width=0.24\textwidth]{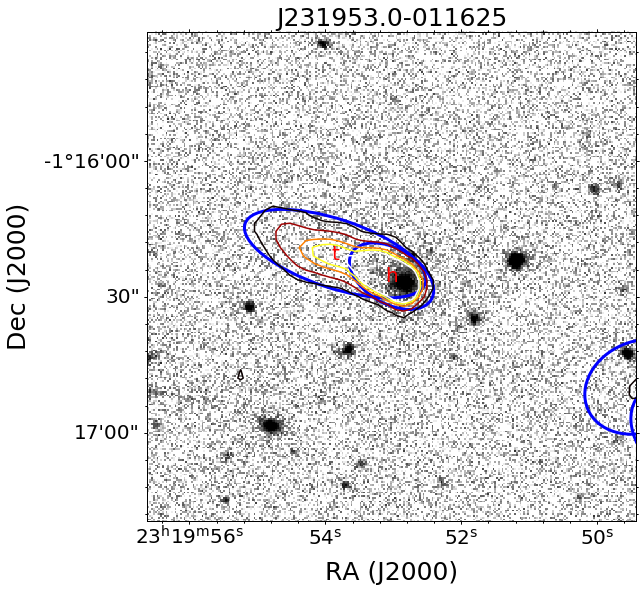} &
\includegraphics[width=0.24\textwidth]{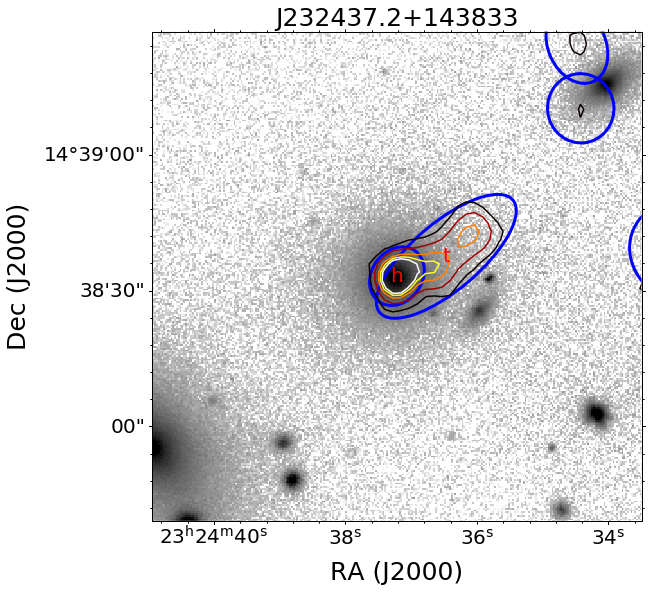} &
\includegraphics[width=0.24\textwidth]{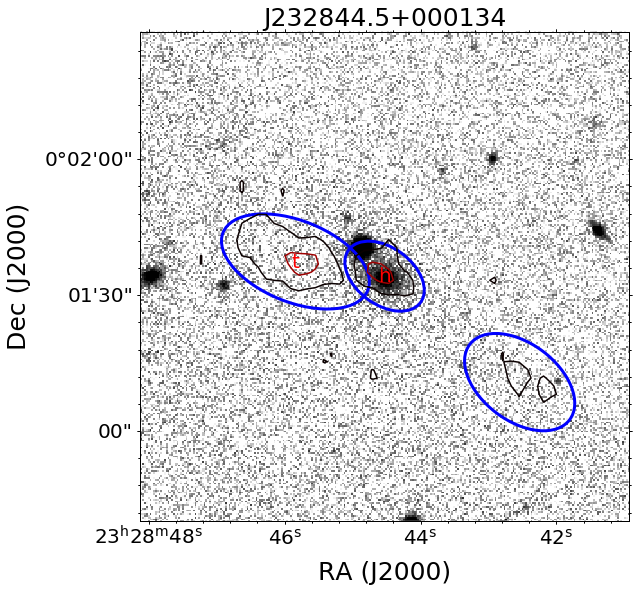} \\
61 & 62 & 63 &64\\
\includegraphics[width=0.24\textwidth]{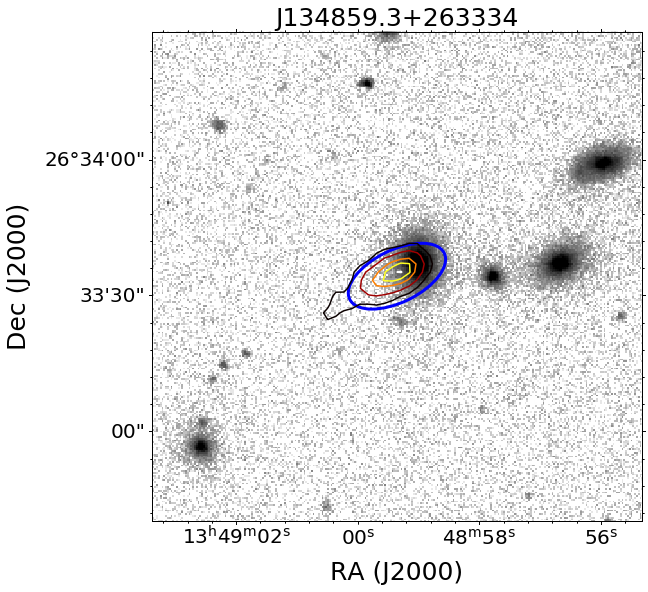} &
\includegraphics[width=0.24\textwidth]{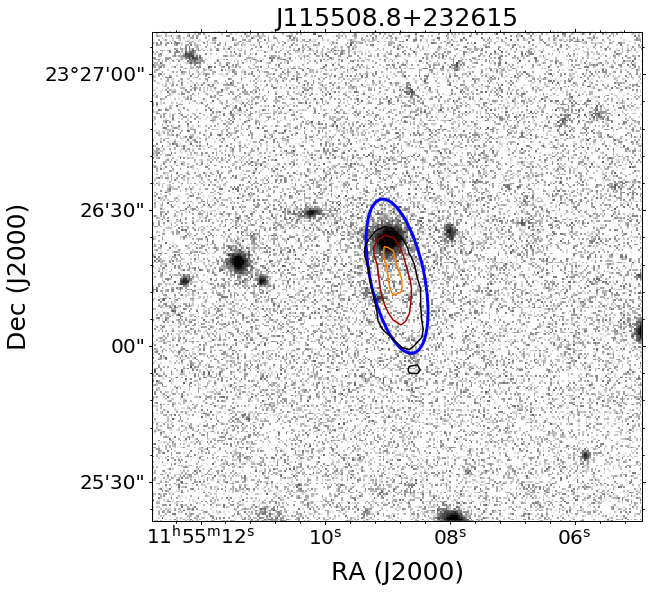} &
\includegraphics[width=0.24\textwidth]{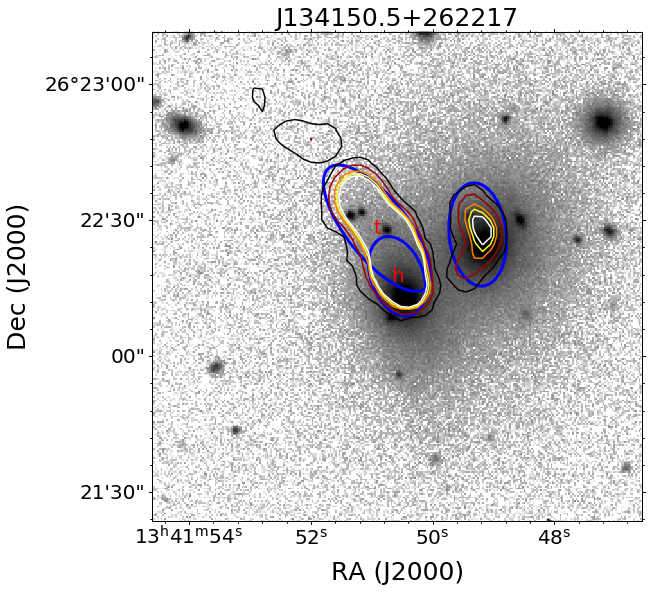} &
\includegraphics[width=0.24\textwidth]{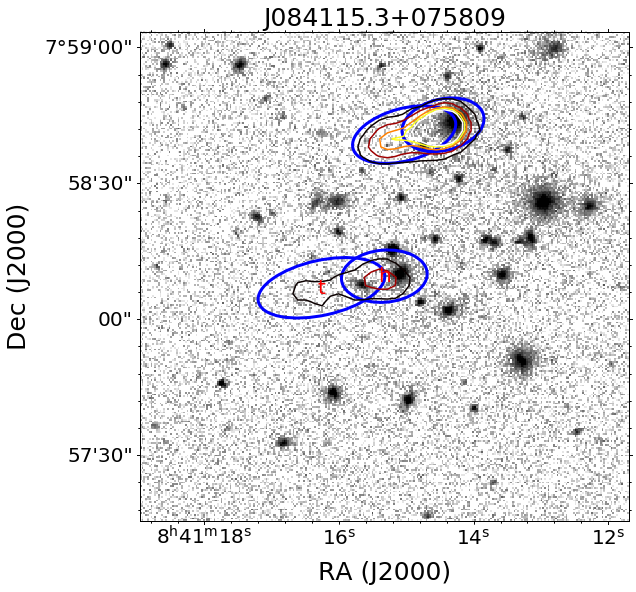} \\
65 & 66 & 67 &68\\
\includegraphics[width=0.24\textwidth]{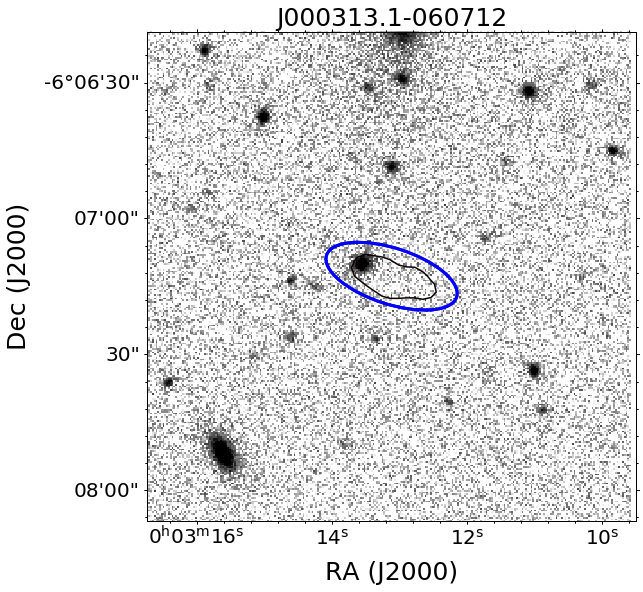} \\
69 \\
\hline
\caption{Identification charts of all 69 OHT candidates. The radio contour maps overlap on top of SDSS r-band images. All images are in the same size 1.8$^\prime$ $\times$ 1.8$^\prime$. The radio contours use the "hot" colormap. The title of each images is the FIRST ID of the ``head" in the center. The blue ellipses on the map represent source shapes from the FIRST catalog. The recognized heads and tails are labelled with the small red ``h" and ``t" correspondingly. }
\end{longtable}
\bibliography{ref} 

\begin{thebibliography}{}
\expandafter\ifx\csname natexlab\endcsname\relax\def\natexlab#1{#1}\fi

\bibitem[{{Abell} {et~al.}(1989){Abell}, {Corwin}, \&
  {Olowin}}]{1989ApJS...70....1A}
{Abell}, G.~O., {Corwin}, Harold~G., J., \& {Olowin}, R.~P. 1989, \apjs, 70, 1

\bibitem[{{Alam} {et~al.}(2015){Alam}, {Albareti}, {Allende Prieto}, {Anders},
  {Anderson}, {Anderton}, {Andrews}, {Armengaud}, {Aubourg}, {Bailey}, {Basu},
  {Bautista}, {Beaton}, {Beers}, {Bender}, {Berlind}, {Beutler}, {Bhardwaj},
  {Bird}, {Bizyaev}, {Blake}, {Blanton}, {Blomqvist}, {Bochanski}, {Bolton},
  {Bovy}, {Shelden Bradley}, {Brandt}, {Brauer}, {Brinkmann}, {Brown},
  {Brownstein}, {Burden}, {Burtin}, {Busca}, {Cai}, {Capozzi}, {Carnero
  Rosell}, {Carr}, {Carrera}, {Chambers}, {Chaplin}, {Chen}, {Chiappini},
  {Chojnowski}, {Chuang}, {Clerc}, {Comparat}, {Covey}, {Croft}, {Cuesta},
  {Cunha}, {da Costa}, {Da Rio}, {Davenport}, {Dawson}, {De Lee}, {Delubac},
  {Deshpande}, {Dhital}, {Dutra-Ferreira}, {Dwelly}, {Ealet}, {Ebelke},
  {Edmondson}, {Eisenstein}, {Ellsworth}, {Elsworth}, {Epstein}, {Eracleous},
  {Escoffier}, {Esposito}, {Evans}, {Fan}, {Fern{\'a}ndez-Alvar}, {Feuillet},
  {Filiz Ak}, {Finley}, {Finoguenov}, {Flaherty}, {Fleming}, {Font-Ribera},
  {Foster}, {Frinchaboy}, {Galbraith-Frew}, {Garc{\'\i}a},
  {Garc{\'\i}a-Hern{\'a}ndez}, {Garc{\'\i}a P{\'e}rez}, {Gaulme}, {Ge},
  {G{\'e}nova-Santos}, {Georgakakis}, {Ghezzi}, {Gillespie}, {Girardi},
  {Goddard}, {Gontcho}, {Gonz{\'a}lez Hern{\'a}ndez}, {Grebel}, {Green},
  {Grieb}, {Grieves}, {Gunn}, {Guo}, {Harding}, {Hasselquist}, {Hawley},
  {Hayden}, {Hearty}, {Hekker}, {Ho}, {Hogg}, {Holley-Bockelmann}, {Holtzman},
  {Honscheid}, {Huber}, {Huehnerhoff}, {Ivans}, {Jiang}, {Johnson},
  {Kinemuchi}, {Kirkby}, {Kitaura}, {Klaene}, {Knapp}, {Kneib}, {Koenig},
  {Lam}, {Lan}, {Lang}, {Laurent}, {Le Goff}, {Leauthaud}, {Lee}, {Lee},
  {Licquia}, {Liu}, {Long}, {L{\'o}pez-Corredoira}, {Lorenzo-Oliveira},
  {Lucatello}, {Lundgren}, {Lupton}, {Mack}, {Mahadevan}, {Maia}, {Majewski},
  {Malanushenko}, {Malanushenko}, {Manchado}, {Manera}, {Mao}, {Maraston},
  {Marchwinski}, {Margala}, {Martell}, {Martig}, {Masters}, {Mathur},
  {McBride}, {McGehee}, {McGreer}, {McMahon}, {M{\'e}nard}, {Menzel},
  {Merloni}, {M{\'e}sz{\'a}ros}, {Miller}, {Miralda-Escud{\'e}}, {Miyatake},
  {Montero-Dorta}, {More}, {Morganson}, {Morice-Atkinson}, {Morrison},
  {Mosser}, {Muna}, {Myers}, {Nand ra}, {Newman}, {Neyrinck}, {Nguyen},
  {Nichol}, {Nidever}, {Noterdaeme}, {Nuza}, {O'Connell}, {O'Connell},
  {O'Connell}, {Ogando}, {Olmstead}, {Oravetz}, {Oravetz}, {Osumi}, {Owen},
  {Padgett}, {Padmanabhan}, {Paegert}, {Palanque-Delabrouille}, {Pan},
  {Parejko}, {P{\^a}ris}, {Park}, {Pattarakijwanich}, {Pellejero-Ibanez},
  {Pepper}, {Percival}, {P{\'e}rez-Fournon}, {Ṕrez-Ra`fols}, {Petitjean},
  {Pieri}, {Pinsonneault}, {Porto de Mello}, {Prada}, {Prakash},
  {Price-Whelan}, {Protopapas}, {Raddick}, {Rahman}, {Reid}, {Rich}, {Rix},
  {Robin}, {Rockosi}, {Rodrigues}, {Rodr{\'\i}guez-Torres}, {Roe}, {Ross},
  {Ross}, {Rossi}, {Ruan}, {Rubi{\~n}o-Mart{\'\i}n}, {Rykoff},
  {Salazar-Albornoz}, {Salvato}, {Samushia}, {S{\'a}nchez}, {Santiago},
  {Sayres}, {Schiavon}, {Schlegel}, {Schmidt}, {Schneider}, {Schultheis},
  {Schwope}, {Sc{\'o}ccola}, {Scott}, {Sellgren}, {Seo}, {Serenelli}, {Shane},
  {Shen}, {Shetrone}, {Shu}, {Silva Aguirre}, {Sivarani}, {Skrutskie},
  {Slosar}, {Smith}, {Sobreira}, {Souto}, {Stassun}, {Steinmetz}, {Stello},
  {Strauss}, {Streblyanska}, {Suzuki}, {Swanson}, {Tan}, {Tayar}, {Terrien},
  {Thakar}, {Thomas}, {Thomas}, {Thompson}, {Tinker}, {Tojeiro}, {Troup},
  {Vargas-Maga{\~n}a}, {Vazquez}, {Verde}, {Viel}, {Vogt}, {Wake}, {Wang},
  {Weaver}, {Weinberg}, {Weiner}, {White}, {Wilson}, {Wisniewski},
  {Wood-Vasey}, {Ye`che}, {York}, {Zakamska}, {Zamora}, {Zasowski}, {Zehavi},
  {Zhao}, {Zheng}, {Zhou}, {Zhou}, {Zou}, \& {Zhu}}]{2015ApJS..219...12A}
{Alam}, S., {Albareti}, F.~D., {Allende Prieto}, C., {et~al.} 2015, \apjs, 219,
  12

\bibitem[{{Banfield} {et~al.}(2015){Banfield}, {Wong}, {Willett}, {Norris},
  {Rudnick}, {Shabala}, {Simmons}, {Snyder}, {Garon}, {Seymour}, {Middelberg},
  {Andernach}, {Lintott}, {Jacob}, {Kapi{\'n}ska}, {Mao}, {Masters}, {Jarvis},
  {Schawinski}, {Paget}, {Simpson}, {Kl{\"o}ckner}, {Bamford}, {Burchell},
  {Chow}, {Cotter}, {Fortson}, {Heywood}, {Jones}, {Kaviraj},
  {L{\'o}pez-S{\'a}nchez}, {Maksym}, {Polsterer}, {Borden}, {Hollow}, \&
  {Whyte}}]{2015MNRAS.453.2326B}
{Banfield}, J.~K., {Wong}, O.~I., {Willett}, K.~W., {et~al.} 2015, \mnras, 453,
  2326

\bibitem[{{Beck} {et~al.}(2016){Beck}, {Dobos}, {Budav{\'a}ri}, {Szalay}, \&
  {Csabai}}]{2016MNRAS.460.1371B}
{Beck}, R., {Dobos}, L., {Budav{\'a}ri}, T., {Szalay}, A.~S., \& {Csabai}, I.
  2016, \mnras, 460, 1371

\bibitem[{{Becker} {et~al.}(1995){Becker}, {White}, \&
  {Helfand}}]{1995ApJ...450..559B}
{Becker}, R.~H., {White}, R.~L., \& {Helfand}, D.~J. 1995, \apj, 450, 559

\bibitem[{{Condon} {et~al.}(1998){Condon}, {Cotton}, {Greisen}, {Yin},
  {Perley}, {Taylor}, \& {Broderick}}]{1998AJ....115.1693C}
{Condon}, J.~J., {Cotton}, W.~D., {Greisen}, E.~W., {et~al.} 1998, \aj, 115,
  1693

\bibitem[{{Das} {et~al.}(2010){Das}, {Gerhard}, {Churazov}, \&
  {Zhuravleva}}]{2010MNRAS.409.1362D}
{Das}, P., {Gerhard}, O., {Churazov}, E., \& {Zhuravleva}, I. 2010, \mnras,
  409, 1362

\bibitem[{{Feretti} \& {Giovannini}(2008)}]{2008LNP...740..143F}
{Feretti}, L., \& {Giovannini}, G. 2008, {Clusters of Galaxies in the Radio:
  Relativistic Plasma and ICM/Radio Galaxy Interaction Processes}, ed.
  M.~{Plionis}, O.~{L{\'o}pez-Cruz}, \& D.~{Hughes}, Vol. 740, 24

\bibitem[{{Garon} {et~al.}(2019){Garon}, {Rudnick}, {Wong}, {Jones}, {Kim},
  {Andernach}, {Shabala}, {Kapi{\'n}ska}, {Norris}, {de Gasperin}, {Tate}, \&
  {Tang}}]{2019AJ....157..126G}
{Garon}, A.~F., {Rudnick}, L., {Wong}, O.~I., {et~al.} 2019, \aj, 157, 126

\bibitem[{{Intema} {et~al.}(2017){Intema}, {Jagannathan}, {Mooley}, \&
  {Frail}}]{2017A&A...598A..78I}
{Intema}, H.~T., {Jagannathan}, P., {Mooley}, K.~P., \& {Frail}, D.~A. 2017,
  \aap, 598, A78

\bibitem[{{Johnston} {et~al.}(2007){Johnston}, {Bailes}, {Bartel}, {Baugh},
  {Bietenholz}, {Blake}, {Braun}, {Brown}, {Chatterjee}, {Darling}, {Deller},
  {Dodson}, {Edwards}, {Ekers}, {Ellingsen}, {Feain}, {Gaensler}, {Haverkorn},
  {Hobbs}, {Hopkins}, {Jackson}, {James}, {Joncas}, {Kaspi}, {Kilborn},
  {Koribalski}, {Kothes}, {Landecker}, {Lenc}, {Lovell}, {Macquart},
  {Manchester}, {Matthews}, {McClure-Griffiths}, {Norris}, {Pen}, {Phillips},
  {Power}, {Protheroe}, {Sadler}, {Schmidt}, {Stairs}, {Staveley-Smith},
  {Stil}, {Taylor}, {Tingay}, {Tzioumis}, {Walker}, {Wall}, \&
  {Wolleben}}]{2007PASA...24..174J}
{Johnston}, S., {Bailes}, M., {Bartel}, N., {et~al.} 2007, \pasa, 24, 174

\bibitem[{{Jones} \& {McAdam}(1996)}]{1996MNRAS.282..137J}
{Jones}, P.~A., \& {McAdam}, W.~B. 1996, \mnras, 282, 137

\bibitem[{{Lacy} {et~al.}(2020){Lacy}, {Baum}, {Chandler}, {Chatterjee},
  {Clarke}, {Deustua}, {English}, {Farnes}, {Gaensler}, {Gugliucci},
  {Hallinan}, {Kent}, {Kimball}, {Law}, {Lazio}, {Marvil}, {Mao}, {Medlin},
  {Mooley}, {Murphy}, {Myers}, {Osten}, {Richards}, {Rosolowsky}, {Rudnick},
  {Schinzel}, {Sivakoff}, {Sjouwerman}, {Taylor}, {White}, {Wrobel},
  {Andernach}, {Beasley}, {Berger}, {Bhatnager}, {Birkinshaw}, {Bower},
  {Brandt}, {Brown}, {Burke-Spolaor}, {Butler}, {Comerford}, {Demorest}, {Fu},
  {Giacintucci}, {Golap}, {G{\"u}th}, {Hales}, {Hiriart}, {Hodge}, {Horesh},
  {Ivezi{\'c}}, {Jarvis}, {Kamble}, {Kassim}, {Liu}, {Loinard}, {Lyons},
  {Masters}, {Mezcua}, {Moellenbrock}, {Mroczkowski}, {Nyland}, {O'Dea},
  {O'Sullivan}, {Peters}, {Radford}, {Rao}, {Robnett}, {Salcido}, {Shen},
  {Sobotka}, {Witz}, {Vaccari}, {van Weeren}, {Vargas}, {Williams}, \&
  {Yoon}}]{2020PASP..132c5001L}
{Lacy}, M., {Baum}, S.~A., {Chandler}, C.~J., {et~al.} 2020, \pasp, 132, 035001

\bibitem[{{Laing} \& {Bridle}(2002)}]{2002MNRAS.336..328L}
{Laing}, R.~A., \& {Bridle}, A.~H. 2002, \mnras, 336, 328

\bibitem[{{Mao} {et~al.}(2009){Mao}, {Johnston-Hollitt}, {Stevens}, \&
  {Wotherspoon}}]{2009MNRAS.392.1070M}
{Mao}, M.~Y., {Johnston-Hollitt}, M., {Stevens}, J.~B., \& {Wotherspoon}, S.~J.
  2009, \mnras, 392, 1070

\bibitem[{{Mazzarella} \& {NED Team}(2017)}]{2017IAUS..325..379M}
{Mazzarella}, J.~M., \& {NED Team}. 2017, in Astroinformatics, ed.
  M.~{Brescia}, S.~G. {Djorgovski}, E.~D. {Feigelson}, G.~{Longo}, \&
  S.~{Cavuoti}, Vol. 325, 379--384

\bibitem[{{McConnachie} {et~al.}(2009){McConnachie}, {Patton}, {Ellison}, \&
  {Simard}}]{2009MNRAS.395..255M}
{McConnachie}, A.~W., {Patton}, D.~R., {Ellison}, S.~L., \& {Simard}, L. 2009,
  \mnras, 395, 255

\bibitem[{{Mendel} {et~al.}(2011){Mendel}, {Ellison}, {Simard}, {Patton}, \&
  {McConnachie}}]{2011MNRAS.418.1409M}
{Mendel}, J.~T., {Ellison}, S.~L., {Simard}, L., {Patton}, D.~R., \&
  {McConnachie}, A.~W. 2011, \mnras, 418, 1409

\bibitem[{{Miley}(1980)}]{1980ARA&A..18..165M}
{Miley}, G. 1980, \araa, 18, 165

\bibitem[{{Miley} {et~al.}(1972){Miley}, {Perola}, {van der Kruit}, \& {van der
  Laan}}]{1972Natur.237..269M}
{Miley}, G.~K., {Perola}, G.~C., {van der Kruit}, P.~C., \& {van der Laan}, H.
  1972, \nat, 237, 269

\bibitem[{{Miraghaei} \& {Best}(2017)}]{2017MNRAS.466.4346M}
{Miraghaei}, H., \& {Best}, P.~N. 2017, \mnras, 466, 4346

\bibitem[{{Mohan} \& {Rafferty}(2015)}]{2015ascl.soft02007M}
{Mohan}, N., \& {Rafferty}, D. 2015, {PyBDSF: Python Blob Detection and Source
  Finder}, , , ascl:1502.007

\bibitem[{{O'Neill} {et~al.}(2019){O'Neill}, {Jones}, {Nolting}, \&
  {Mendygral}}]{2019ApJ...887...26O}
{O'Neill}, B.~J., {Jones}, T.~W., {Nolting}, C., \& {Mendygral}, P.~J. 2019,
  \apj, 887, 26

\bibitem[{{Pratley} {et~al.}(2013){Pratley}, {Johnston-Hollitt}, {Dehghan}, \&
  {Sun}}]{2013MNRAS.432..243P}
{Pratley}, L., {Johnston-Hollitt}, M., {Dehghan}, S., \& {Sun}, M. 2013,
  \mnras, 432, 243

\bibitem[{{Proctor}(2011)}]{2011ApJS..194...31P}
{Proctor}, D.~D. 2011, \apjs, 194, 31

\bibitem[{{Ryle} \& {Windram}(1968)}]{1968MNRAS.138....1R}
{Ryle}, M., \& {Windram}, M.~D. 1968, \mnras, 138, 1

\bibitem[{{Savini} {et~al.}(2019){Savini}, {Bonafede}, {Br{\"u}ggen},
  {Rafferty}, {Shimwell}, {Botteon}, {Brunetti}, {Intema}, {Wilber}, {Cassano},
  {Vazza}, {van Weeren}, {Cuciti}, {De Gasperin}, {R{\"o}ttgering}, {Sommer},
  {B{\^\i}rzan}, \& {Drabent}}]{2019A&A...622A..24S}
{Savini}, F., {Bonafede}, A., {Br{\"u}ggen}, M., {et~al.} 2019, \aap, 622, A24

\bibitem[{{Shimwell} {et~al.}(2017){Shimwell}, {R{\"o}ttgering}, {Best},
  {Williams}, {Dijkema}, {de Gasperin}, {Hardcastle}, {Heald}, {Hoang},
  {Horneffer}, {Intema}, {Mahony}, {Mandal}, {Mechev}, {Morabito}, {Oonk},
  {Rafferty}, {Retana-Montenegro}, {Sabater}, {Tasse}, {van Weeren},
  {Br{\"u}ggen}, {Brunetti}, {Chy{\.z}y}, {Conway}, {Haverkorn}, {Jackson},
  {Jarvis}, {McKean}, {Miley}, {Morganti}, {White}, {Wise}, {van Bemmel},
  {Beck}, {Brienza}, {Bonafede}, {Calistro Rivera}, {Cassano}, {Clarke},
  {Cseh}, {Deller}, {Drabent}, {van Driel}, {Engels}, {Falcke}, {Ferrari},
  {Fr{\"o}hlich}, {Garrett}, {Harwood}, {Heesen}, {Hoeft}, {Horellou},
  {Israel}, {Kapi{\'n}ska}, {Kunert-Bajraszewska}, {McKay}, {Mohan},
  {Orr{\'u}}, {Pizzo}, {Prandoni}, {Schwarz}, {Shulevski}, {Sipior}, {Smith},
  {Sridhar}, {Steinmetz}, {Stroe}, {Varenius}, {van der Werf}, {Zensus}, \&
  {Zwart}}]{2017A&A...598A.104S}
{Shimwell}, T.~W., {R{\"o}ttgering}, H.~J.~A., {Best}, P.~N., {et~al.} 2017,
  \aap, 598, A104

\bibitem[{{Shimwell} {et~al.}(2019){Shimwell}, {Tasse}, {Hardcastle}, {Mechev},
  {Williams}, {Best}, {R{\"o}ttgering}, {Callingham}, {Dijkema}, {de Gasperin},
  {Hoang}, {Hugo}, {Mirmont}, {Oonk}, {Prandoni}, {Rafferty}, {Sabater},
  {Smirnov}, {van Weeren}, {White}, {Atemkeng}, {Bester}, {Bonnassieux},
  {Br{\"u}ggen}, {Brunetti}, {Chy{\.z}y}, {Cochrane}, {Conway}, {Croston},
  {Danezi}, {Duncan}, {Haverkorn}, {Heald}, {Iacobelli}, {Intema}, {Jackson},
  {Jamrozy}, {Jarvis}, {Lakhoo}, {Mevius}, {Miley}, {Morabito}, {Morganti},
  {Nisbet}, {Orr{\'u}}, {Perkins}, {Pizzo}, {Schrijvers}, {Smith}, {Vermeulen},
  {Wise}, {Alegre}, {Bacon}, {van Bemmel}, {Beswick}, {Bonafede}, {Botteon},
  {Bourke}, {Brienza}, {Calistro Rivera}, {Cassano}, {Clarke}, {Conselice},
  {Dettmar}, {Drabent}, {Dumba}, {Emig}, {En{\ss}lin}, {Ferrari}, {Garrett},
  {G{\'e}nova-Santos}, {Goyal}, {G{\"u}rkan}, {Hale}, {Harwood}, {Heesen},
  {Hoeft}, {Horellou}, {Jackson}, {Kokotanekov}, {Kondapally},
  {Kunert-Bajraszewska}, {Mahatma}, {Mahony}, {Mandal}, {McKean}, {Merloni},
  {Mingo}, {Miskolczi}, {Mooney}, {Nikiel-Wroczy{\'n}ski}, {O'Sullivan},
  {Quinn}, {Reich}, {Roskowi{\'n}ski}, {Rowlinson}, {Savini}, {Saxena},
  {Schwarz}, {Shulevski}, {Sridhar}, {Stacey}, {Urquhart}, {van der Wiel},
  {Varenius}, {Webster}, \& {Wilber}}]{2019A&A...622A...1S}
{Shimwell}, T.~W., {Tasse}, C., {Hardcastle}, M.~J., {et~al.} 2019, \aap, 622,
  A1

\bibitem[{{Smith} {et~al.}(2012){Smith}, {Hopkins}, {Hunstead}, \&
  {Pimbblet}}]{2012MNRAS.422...25S}
{Smith}, A.~G., {Hopkins}, A.~M., {Hunstead}, R.~W., \& {Pimbblet}, K.~A. 2012,
  \mnras, 422, 25

\bibitem[{{Sparks} {et~al.}(1992){Sparks}, {Fraix-Burnet}, {Macchetto}, \&
  {Owen}}]{1992Natur.355..804S}
{Sparks}, W.~B., {Fraix-Burnet}, D., {Macchetto}, F., \& {Owen}, F.~N. 1992,
  \nat, 355, 804

\bibitem[{{Srivastava} \& {Singal}(2020)}]{2020MNRAS.493.3811S}
{Srivastava}, S., \& {Singal}, A.~K. 2020, \mnras, 493, 3811

\bibitem[{{Terni de Gregory} {et~al.}(2017){Terni de Gregory}, {Feretti},
  {Giovannini}, {Govoni}, {Murgia}, {Perley}, \& {Vacca}}]{2017A&A...608A..58T}
{Terni de Gregory}, B., {Feretti}, L., {Giovannini}, G., {et~al.} 2017, \aap,
  608, A58

\bibitem[{{Wen} \& {Han}(2015)}]{2015ApJ...807..178W}
{Wen}, Z.~L., \& {Han}, J.~L. 2015, \apj, 807, 178

\bibitem[{{White} {et~al.}(1997){White}, {Becker}, {Helfand}, \&
  {Gregg}}]{1997ApJ...475..479W}
{White}, R.~L., {Becker}, R.~H., {Helfand}, D.~J., \& {Gregg}, M.~D. 1997,
  \apj, 475, 479

\bibitem[{{Yu} {et~al.}(2018){Yu}, {Tozzi}, {van Weeren}, {Liuzzo},
  {Giovannini}, {Donahue}, {Balestra}, {Rosati}, \&
  {Aravena}}]{2018ApJ...853..100Y}
{Yu}, H., {Tozzi}, P., {van Weeren}, R., {et~al.} 2018, \apj, 853, 100

\end{thebibliography}

\end{document}